%% file: main.tex
\documentclass[twocolumn]{aa}

\usepackage{graphicx}
\usepackage{amsmath,amsfonts,amssymb}
\usepackage{txfonts}
\usepackage{color}
\usepackage{natbib}
\bibpunct{(}{)}{;}{a}{}{,} 
\usepackage{float}
\usepackage{dblfloatfix}
\usepackage{afterpage}
\usepackage{ifthen}
\usepackage[morefloats=12]{morefloats}
\usepackage{placeins}
\usepackage{multicol}
\bibpunct{(}{)}{;}{a}{}{,}
\usepackage[switch]{lineno}
\definecolor{linkcolor}{rgb}{0.6,0,0}
\definecolor{citecolor}{rgb}{0,0,0.75}
\definecolor{urlcolor}{rgb}{0.12,0.46,0.7}
\PassOptionsToPackage{hyphens}{url}
\usepackage{url}
\usepackage[breaklinks, colorlinks, urlcolor=urlcolor,
    linkcolor=linkcolor,citecolor=citecolor,pdfencoding=auto]{hyperref}
\hypersetup{linktocpage}
\usepackage{bold-extra}
\usepackage{capt-of}

\input{Planck}

\def\WMAP{\emph{WMAP}}
\def\WMAPnine{\emph{WMAP9}}
\def\COBE{\emph{COBE}}
\def\wmap{\emph{WMAP}}

\def\Planck{\emph{Planck}}

\def\healpix{\texttt{HEALPix}}
\def\commander{\texttt{Commander}}
\def\commanderone{\texttt{Commander1}}

\def\commanderthree{\texttt{Commander3}}

\def\sroll2{\texttt{SRoll2}}

\renewcommand{\L}[0]{\tens{L}}

\newcommand{\BP}{\textsc{BeyondPlanck}}
\newcommand{\bp}{\textsc{BeyondPlanck}}
\newcommand{\cosmoglobe}{\textsc{Cosmoglobe}}
\newcommand{\Cosmoglobe}{\textsc{Cosmoglobe}}

\newcommand{\K}[0]{\textit K}
\newcommand{\Ka}[0]{\textit{Ka}}
\newcommand{\Q}[0]{\textit Q}
\newcommand{\V}[0]{\textit V}
\newcommand{\W}[0]{\textit W}

\usepackage[T1]{fontenc}

\def\inv{^{-1}}

\begin{document}

\title{\bfseries{\Cosmoglobe\ DR1. III. First full-sky model of polarized synchrotron\\ emission from all \WMAP\ and \Planck\ LFI data}}
\input{authors.tex}
\authorrunning{Watts et al.}
\titlerunning{\cosmoglobe\ Polarized Synchrotron Analysis}

\abstract{
  We present the first model of full-sky polarized synchrotron emission that is derived from all \WMAP\ and \Planck\ LFI frequency maps. The basis of this analysis is the set of end-to-end reprocessed \cosmoglobe\ Data Release 1 sky maps presented in a companion paper, which have significantly lower instrumental systematics than the legacy products from each experiment. We find that the resulting polarized synchrotron amplitude map has an average noise rms of $3.2\,\mathrm{\mu K}$ at 30\,GHz and $2^{\circ}$ FWHM, which is 30\,\% lower than the recently released \BP\ model that included only LFI+\WMAP\ \Ka--\V\ data, and 29\,\% lower than the \WMAP\ \K-band map alone. The mean $B$-to-$E$ power spectrum ratio is $0.40\pm0.02$, with amplitudes consistent with those measured previously by  \Planck\ and QUIJOTE. Assuming a power law model for the synchrotron spectral energy distribution, and using the $T$--$T$ plot method, we find a full-sky inverse noise-variance weighted mean of $\beta_{\mathrm{s}}=-3.07\pm0.07$ between \cosmoglobe\ DR1 \K-band and 30\,GHz, in good agreement with previous estimates. In summary, the novel \cosmoglobe\ DR1 synchrotron model is both more sensitive and systematically cleaner than similar previous models, and it has a more complete error description that is defined by a set of Monte Carlo posterior samples. We believe that these products are preferable over previous \Planck\ and \WMAP\ products for all synchrotron-related scientific applications, including simulation, forecasting and component separation. 
}


\maketitle


\section{Introduction}
\label{sec:introduction}

Understanding the polarization of the cosmic microwave background (CMB) is a primary focus of observational cosmology in the coming decades. There has been phenomonal observational success over the past few decades, from the satellite-based \COBE\ \citep{smoot:1992,mather:1994,hauser:1998}, \textit{Wilkinson Microwave Anisotropy Probe} (\WMAP) \citep{bennett2012}, and \Planck\ \citep{planck2016-l01} experiments, to the many sub-orbital experiments including for instance ACT \citep{actDR6_lensing}, BICEP/\textit{Keck} \citep{bicep2021}, CLASS \citep{eimer2023}, QUIJOTE \citep{QUIJOTE_IV}, SPIDER \citep{spider21}, SPT \citep{carlstrom:2011} and many others. Future experiments, including the \textit{LiteBIRD} satellite \citep{litebird2022},  Simons Observatory \citep{SO2019} and CMB-S4 \citep{cmbs4},  will create the most sensitive maps of the polarized sky yet, resulting in stringent constraints on primordial gravitational waves.

The dominant astrophysical sources for polarized radiation in the microwave frequency range is synchrotron emission from relativistic electrons moving through the magnetic field of the Milky Way, and thermal dust emission from vibrating dust grains. In order to map the valuable CMB fluctuations, both of these contaminants must be characterized and subtracted at high precision. Indeed, as shown by both \Planck\ and BICEP2/\textit{Keck}, uncertainties on polarization-based cosmological constraints have been limited not only by instrumental sensitivity, but by incomplete knowledge of the sky itself. Uncertainty in the sky model has been mitigated by designing experiments with broad frequency coverage, such as \WMAP\ \citep{bennett2012} and \Planck\ \citep{planck2016-l01}, or analyzing maps from different experiments jointly (e.g., \citealp{dmr}, \citealp{bennett2012}, \citealp{planck2014-a12}, and \citealp{pb2015}, among many others).
A major impediment to joint analysis is the difficulty of combining individually analyzed data with different survey strategies and incompletely characterized systematics. In order to maximize scientific throughput, one must either design an experiment that can characterize every relevant observable on its own, or jointly analyze different datasets in the same joint framework. In practice, joint analysis has therefore  until now usually been performed by combining datasets at the likelihood level.

The \bp\ project performed a conceptually different form of joint analysis by combining the \Planck\ Low Frequency Instrument (LFI) time-ordered data (TOD) \citep{bp01} with external pixel-based data directly in the same analysis pipeline, including the Haslam 408\,MHz map \citep{haslam1982}, \WMAP\ \Ka--\V\ bands, \Planck\ 353\,GHz in polarization, and \Planck\ 857\,GHz in intensity. \citet{planck2016-l02} found that the LFI detector gain solution depended on the assumed polarization and intensity of the sky. To break this circular dependency, the \bp\ framework solved for the intrinsic sky signal and instrumental parameters iteratively, providing an accurate model of the entire system with full error propagation. By leveraging the external \WMAP, Haslam, and \Planck\ HFI  data, \bp\ was able to create \Planck\ LFI maps of cosmological quality at all frequency channels, while simultaneously generating a robust model of the foreground sky.

The \cosmoglobe\ project\footnote{\url{cosmoglobe.uio.no}} has now generalized the \bp\ analysis by also processing  \WMAP\ \K--\W-band as time-ordered data. This framework, fully described by \citet{bp17} and \cite{watts2023_dr1},
not only improves the quality of the \WMAP\ maps themselves, but provides the most robust full-sky model of low frequency polarized emission available to date. In particular, poorly-measured modes due to transmission imbalance \citep{jarosik2007} have been effectively marginalized over, in large part due to the use of a global sky model that was not available for the fiducial \WMAP\ analyses. The removal of poorly measured modes alone will improve Galactic modelling for years to come, as the current state-of-the-art models use \WMAP\ \K-band (23\,GHz) maps as polarized synchrotron templates for modeling of synchrotron without any mitigation of the poorly-measured modes \citep{delabrouille2012,pysm2,pysm3}.

In order to make use of the synchrotron model, we require robust estimates of the spectral behavior. Fundamental properties of synchrotron emission, such as the spectral index as a function of position on the sky, spatial decorrelation, and even the functional form of the spectral energy distribution (SED), are yet to be fully characterized observationally. Modern attempts have been frustrated by the lack of high signal-to-noise data; \citet{deBelsunce:2022}, for example, find spectral indices ranging from $-5$ to $-1$, discrepant with predictions from direct measurements of the cosmic ray energy spectrum of 2--3 \citep{rybicki,orlando2013,neronov2017}, largely due to fitting too many parameters to data with too low signal-to-noise ratios. Such analyses can be greatly improved by combining complementary data sets.

While the combination of as many datasets as possible would help to constrain polarized synchrotron's large-scale properties, long-standing discrepancies between existing datasets complicate this. As shown by, for example, \citet{planck2014-a12} and \citet{weiland:2018}, there are discrepancies in the polarization measurements of \WMAP\ and \Planck, partially due to known instrumental effects \citep{bennett2012,planck2016-l02}. \citet{bp01} largely resolved these issues for LFI, while \citet{bp17} removed the poorly measured modes in \WMAP. In this first \cosmoglobe\ data release, of which this paper is a part, \citet{watts2023_dr1} demonstrate that the polarized maps resulting from these two experiments are consistent at the noise level after recalibrating both experiments jointly. At the same time, the new maps must be validated by a variety of analysis methods. As noted in, for example, \citet{weiland:2022}, simply using a different analysis pipeline can result in different estimates of the underlying frequency map. As such, we take care to appopriately marginalize over instrumental effects in our spectral analysis, and attempt to find the physical reason for the differences between different pipelines.

With these longstanding discrepancies addressed, the main goal of the current paper is to derive the highest signal-to-noise ratio model of polarized synchrotron emission from \Planck\ and \WMAP\ data published to date, including both amplitude and spectral index parameters. This work is organized as follows. In Sect.~\ref{sec:data}, we provide a brief overview of the data products and the \cosmoglobe\ data processing. Section~\ref{sec:pol_amp} discusses the polarization amplitude of the \cosmoglobe\ synchrotron model, its overall uncertainty, and the power spectrum properties compared to external products. Section~\ref{sec:specvar} estimates the synchrotron spectral index using both the $T$--$T$ plot method and Gibbs sampling using \commander, followed by discussions and conclusions in Sect.~\ref{sec:conclusion}.



\section{Data products}
\label{sec:data}

\begin{figure*}
	\centering
	\includegraphics[width=0.2\textwidth]{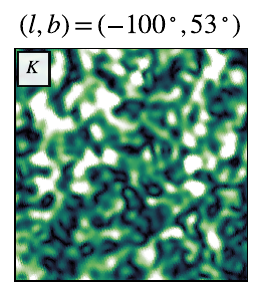}
	\includegraphics[width=0.35\textwidth]{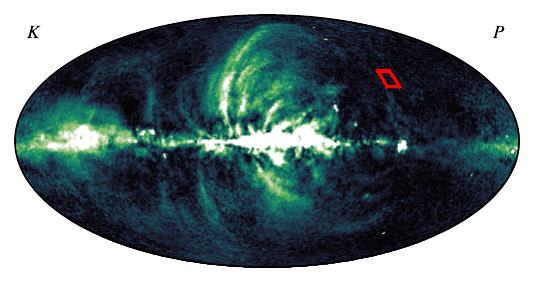}
	\includegraphics[width=0.35\textwidth]{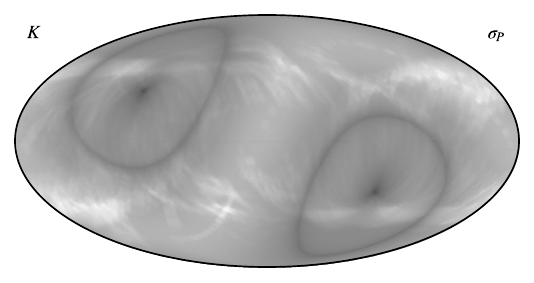}\\
	\includegraphics[width=0.2\textwidth]{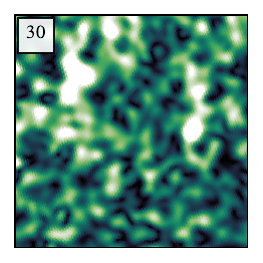}
	\includegraphics[width=0.35\textwidth]{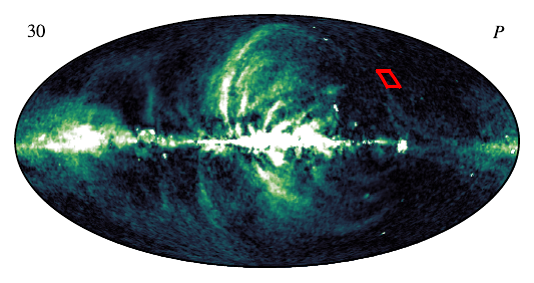}
	\includegraphics[width=0.35\textwidth]{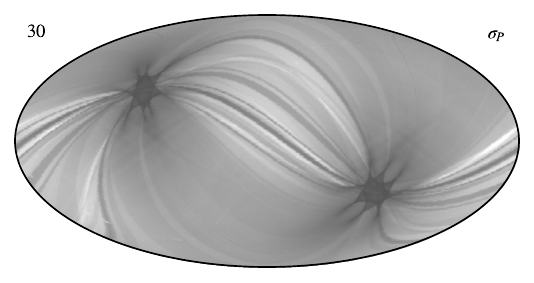}\\
	\includegraphics[width=0.212\textwidth]{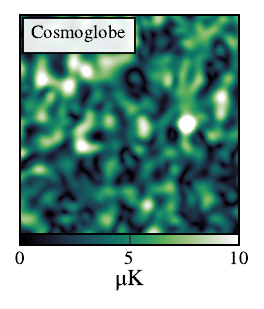}
	\includegraphics[width=0.35\textwidth]{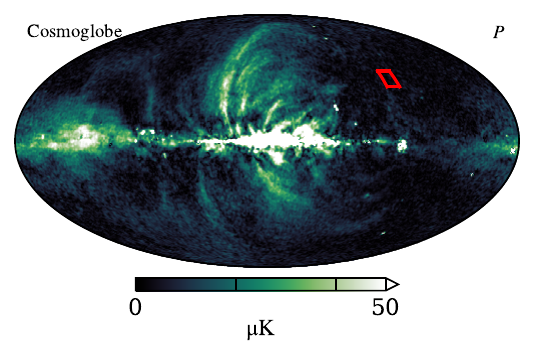}
	\includegraphics[width=0.35\textwidth]{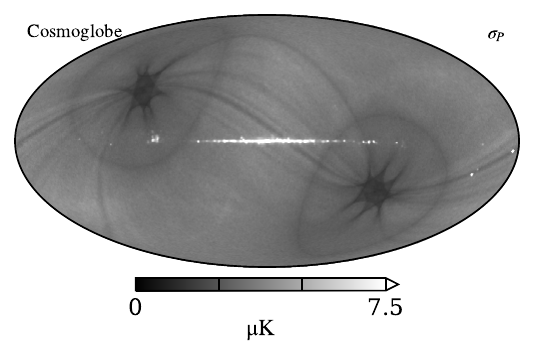}
	\caption{
		Polarized intensity and white noise levels of \textit{(top):} \WMAP\ \K-band, \textit{(middle):} \Planck\ 30\,GHz, and \textit{(bottom):} synchrotron amplitude from the \cosmoglobe\ Gibbs chain, all evaluated at 30\,GHz with a resolution of $72\arcm$. The leftmost column is a $10^\circ$ width square centered on the low signal-to-noise region highlighted by a red square in the middle column's panels; the middle column shows the polarized total amplitude, $P$; and the rightmost column shows the rms noise for the \K-band and 30\,GHz frequency maps and the \cosmoglobe\ DR1 synchrotron amplitude map.
		}
       \label{fig:synch_polint}
\end{figure*}

The data products used in this paper are the \WMAP\ and \Planck\ LFI maps, with most of the statistical weight coming from 23--33\,GHz, or the \WMAP\ \K\ and \Ka\ bands and LFI's 30\,GHz band. These frequencies are low enough that we can treat them as synchrotron tracers and hence ignore thermal dust and CMB emission, but not so low that we need to take into account effects like Faraday rotation, such as in S-PASS data \citep{krachmalnicoff2018,fuskeland:2019}. We describe the legacy data products and the data products from \cosmoglobe\ DR1 in the following subsections.

\subsection{\WMAP\ and \Planck\ legacy products}
\label{sec:wmap_data}

\WMAP\ was a NASA-funded satellite mission that observed from August 2001 to August 2010, designed to characterize the microwave sky well enough to measure the primary CMB anisotropies across the full sky down to a resolution of 13\arcm\ FWHM. Using a differential scanning strategy inspired by \COBE/DMR,
\WMAP\ produced maps of the sky at 23 (\K), 33 (\Ka), 41 (\Q), 61 (\V), and 94\,GHz (\W) in both polarization and total intensity \citep{bennett2012}, with angular resolutions of 53\arcm\ at 23\,GHz to 13\arcm\ at 94\,\GHz. 
The maps are available on the LAMBDA website.\footnote{\url{https://lambda.gsfc.nasa.gov/product/wmap/dr5/m_products.html}} 

The \Planck\ Low Frequency Instrument (LFI) produced  30, 44, and 70 GHz maps in both intensity and polarization, while the High Frequency Instrument (HFI) produced 100, 143, 217, 353\,GHz maps in polarization and intensity, and 545 and 857\,GHz maps in intensity alone. The LFI data have lower white noise and higher angular resolutions (of 30\arcm, 20\arcm, and 13\arcm\ FWHM for 30, 44, and 70\,GHz) than \WMAP. In contrast to \WMAP, the LFI measurements used a single horn, and the \Planck\ scanning strategy followed rings closely aligned with ecliptic meridians. The \Planck\ legacy datasets, PR3 \citep{planck2016-l01} and PR4 \citep{planck2020-LVII}, are both publicly available on the \Planck\ Legacy Archive (PLA).\footnote{\url{https://pla.esac.esa.int/}}

\subsection{\Cosmoglobe\ products}
\label{sec:cosmoglobe_data}

A main goal of \Cosmoglobe\ is to perform joint end-to-end analyses on multiple data sets, preferably beginning from raw TOD. An important advantage of such end-to-end processing is that it offers a robust path to breaking internal degeneracies within and between different data sets, in general reducing the magnitude of systematic effects. The analysis described by \citet{bp01} and \citet{watts2023_dr1} were performed on raw TOD, producing cosmological parameters using a Bayesian Gibbs sampler called \commanderthree\ \citep{bp03}. 

In this framework, we produce a full sky model and set of instrumental parameters for each Gibbs sample, and the set of all such samples allows for a thorough characterization of the dependence of low-level instrumental parameters on the sky model. The sky model includes all relevant components in \WMAP\ and LFI's frequency range, specifically the CMB, synchrotron, thermal dust, free-free emission, anomalous microwave emission, and radio point sources, the first three of which we model as polarized. The full products from this analysis and individual maps are available on the \cosmoglobe\ website.\footnote{\href{https://www.cosmoglobe.uio.no/products/cosmoglobe-dr1.html}{\texttt{https://www.cosmoglobe.uio.no/products/\newline cosmoglobe-dr1.html}}}

As described by \citet{watts2023_dr1}, \cosmoglobe\ DR1 includes 500 end-to-end Gibbs samples, and we perform our basic analysis on each of these samples individually, then form posterior summary statistics from the full ensemble. This allows us to fully marginalize over the low-level systematic parameters, quantifying the extent to which instrumental processing propogates to the synchrotron spectral index determination.

\section{Polarized synchrotron amplitude}
\label{sec:pol_amp}

In this section, we give an overview of the polarized synchrotron amplitude properties. In Sec.~\ref{sec:pol_amp_map}, we focus on the map-based properties, and compare with independently processed results in Sec.~\ref{sec:comp_independent}. In Sec.~\ref{sec:powspec},  we evaluate the power spectrum of the polarized synchrotron map and compute the $B$-to-$E$ ratio.

\subsection{Polarization amplitude maps}
\label{sec:pol_amp_map}



We start by comparing the polarized spectral amplitude (as defined by $P=\sqrt{Q^2+U^2}$) from the \cosmoglobe\ synchrotron map with the official \WMAP\ \K-band and \Planck\ 30\,GHz maps in Fig.~\ref{fig:synch_polint}. The noise level of the \cosmoglobe\ synchrotron map is already at a visual level clearly lower than those of the \K-band and 30\,GHz maps. This is particularly visible in the leftmost column of Fig.~\ref{fig:synch_polint}, an inset of a high Galactic latitude, low signal region.

To quantify the noise improvement in the synchrotron maps over pure templates based on either \K-band or 30\,GHz maps, we compare the posterior standard deviation of the \cosmoglobe\ synchrotron map with the frequency maps' rms values.
To do so, it is important to take into account the fact that the \cosmoglobe\ synchrotron map has effectively been convolved with a $30\arcm$ FWHM Gaussian smoothing prior, which suppresses small-scale fluctuations. To compare the \K-band and 30\,GHz noise levels with this product, we therefore simulate noise realizations, smooth with a corresponding beam, and scale the maps to the default reference frequency of 30\,GHz assuming $\beta_\mathrm s=-3.1$, consistent with the post-processed synchrotron map. We display these smoothed rms maps in the right column of Fig.~\ref{fig:synch_polint}.

Doing this, we find that the mean rms for \K-band and 30\,GHz are $4.8\,\mathrm{\mu K}$ and $4.7\,\mathrm{\mu K}$ respectively, compared to the mean value of $3.4\,\mathrm{\mu K}$ for the \cosmoglobe\ DR1 synchrotron map; this is consistent with adding the scaled maps in quadrature.  While this may seem obvious on its face,  the combination of \WMAP\ \K-band and \Planck\ 30\,GHz is not straightforward due to instrumental effects that remain in the maps, notably poorly measured modes in \WMAP\ \citep{bennett2012,weiland:2018} and gain uncertainty in \Planck\ \citep{planck2016-l02}. These effects remain in the official \WMAPnine\ \citep{bennett2012} and \Planck\ PR3/PR4 \citep{planck2016-l02,planck2020-LVII} maps, making combination of these datasets non-ideal for Galactic science and cosmological analyses. However, the end-to-end \bp\ \citep{bp01} products effectively removed the gain uncertainty modes from the \Planck\ LFI maps, and the \cosmoglobe\ DR1 results \citep{watts2023_dr1} are free from the poorly measured modes. As shown in Fig.~46 of \citet{watts2023_dr1}, the \Planck\ 30\,GHz and the \WMAP\ \K-band are consistent with each other at the $10\,\mathrm{\mu K}$ level, which corresponds to white noise. This polarized synchrotron map, derived from consistent datasets, has a white noise level 29\,\% lower than both the \K-band map and the 30\,GHz map.




\begin{figure}
	\centering
	\includegraphics[width=0.45\textwidth]{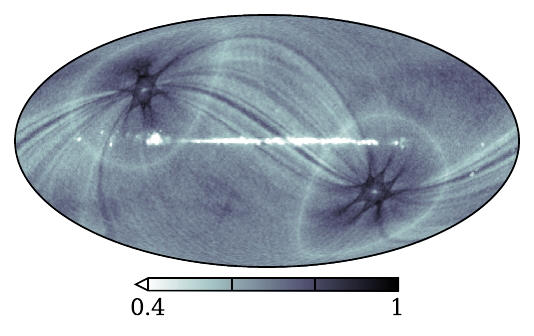}
	\caption{
		Ratio between \cosmoglobe\ and \bp\ polarized synchrotron amplitude marginal posterior deviation per HEALPix $N_{\mathrm{side}}=1024$ pixel.
		}
		\label{fig:rms_ratios}
\end{figure}

In Fig.~\ref{fig:rms_ratios} we plot the ratio between the \cosmoglobe\ and \bp\ \citep{bp01} synchrotron rms maps. Both these products are created using the \commanderthree\ pipeline and similar data selection, but \cosmoglobe\ benefits from additionally using the \K-band data. The \cosmoglobe\ synchrotron map has a mean posterior rms of $3.4\,\mathrm{\mu K}$, nearly a factor of two improvement over the \bp\ map.
As expected, the regions with the lowest ratios correspond to the deep \WMAP\ observations and the Galactic plane, which benefit from the high signal-to-noise of the \K-band data. Regions with nearly identical rms include regions with the highest \Planck\ depth, such as the ecliptic poles, and regions less deeply observed by \WMAP, corresponding to planet crossings and artifacts from the processing mask. Notably, stripes corresponding the \Planck\ scan strategy also show improvement with respect to \bp. This is due to the interaction between the sky model and LFI's instrumental parameters -- with the high signal-to-noise \K-band data, the sky model becomes more stable, and LFI's relative gain solution becomes better determined.

\subsection{Comparison with independent datasets}
\label{sec:comp_independent}

\begin{figure*}
	\begin{center}
	\includegraphics[width=0.45\textwidth]{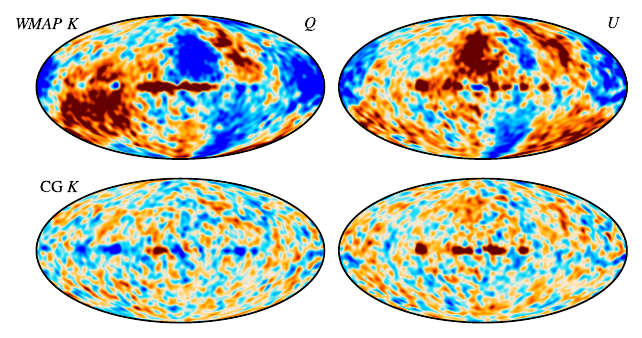}
	\includegraphics[width=0.45\textwidth]{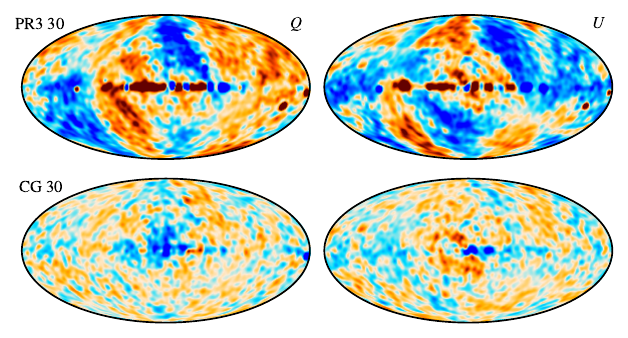}
	\includegraphics[width=0.45\textwidth]{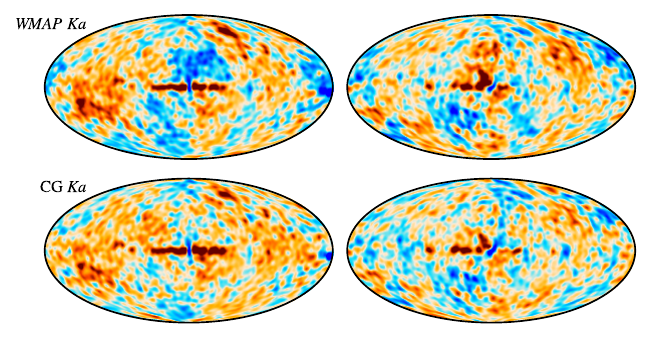}
	\includegraphics[width=0.45\textwidth]{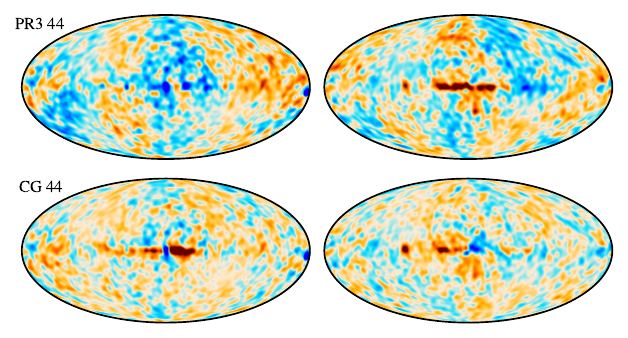}
        \includegraphics[width=0.25\textwidth]{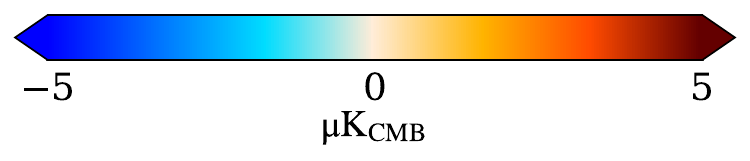}
	\end{center}
	\caption{Frequency map minus sky model with respect to the \cosmoglobe\ DR1 sky model evaluated at $5^\circ$. \cosmoglobe\ maps are labeled CG, \Planck\ 2018 maps are labeled PR3, and the legacy \WMAPnine\ maps are labeled \WMAP. Each of the \WMAPnine, \Planck\ PR3, and \cosmoglobe\ maps have had the \cosmoglobe\ DR1 sky model evaluated at their respective bandpass and resolution subtracted. 
	}
	\label{fig:cg_residuals}
\end{figure*}

Next, we compare the \cosmoglobe\ polarized synchrotron map with the frequency maps produced in the main DR1 chain, paying special attention to maps with the highest polarized synchrotron signal-to-noise ratio; \K, \Ka, 30\,GHz, and 44\,GHz.
Using the polarized synchrotron model generated in the main \cosmoglobe\ DR1 chain, we evaluate the synchrotron emission at each frequency and subtract this from frequency maps produced by various different processing pipelines. We evaluate the \cosmoglobe\ sky model using the \texttt{cosmoglobe} Python package using the full bandpass information of each instrument.\footnote{\url{https://cosmoglobe.readthedocs.io/en/latest/tutorials/skymodel.html}} 


The first and third rows of Fig.~\ref{fig:cg_residuals} show the \WMAPnine\ and PR3 residuals with respect to the \cosmoglobe\ sky model. These residuals match previously-documented observational effects in both experiments. In particular, the \WMAP\ maps show the poorly-measured modes due to transmission imbalance in the differential horns \citep{jarosik2007,bennett2012}, while the PR3  differences are mostly  due to relative gain errors \citep{planck2016-l02,planck2020-LVII}.
In contrast, as seen in the second and fourth rows of Fig.~\ref{fig:cg_residuals}, the synchrotron model matches the \cosmoglobe\ DR1 frequency maps within $5\,\mathrm{\mu K}$ across the sky, with few observational artifacts. Most of the \cosmoglobe\ residuals are associated with the Galactic plane and diffuse structures uncorrelated with the \WMAP\ and \Planck\ observation strategies. The positive excess in the \Ka\ Stokes $Q$ map could be due to unmitigated data processing effects. While a similar large-scale excess can also be seen in the LFI 30 and 44\,GHz Stokes $Q$ maps, the signature strongly resembles the bandpass correction in \WMAP\ \K-band, suggesting incomplete treatment of bandpasses could be contributing to this large-scale signal.
The LFI residuals, while much improved, still show trace residuals, especially near the Galactic center, that are somewhat correlated with the gain correction templates, but not at a level that high Galactic latitude features can be identified. 


\subsection{Power spectra}
\label{sec:powspec}


Next, we consider the angular power spectrum of polarized synchrotron emission evaluated at 30\,GHz. To estimate the power spectra without noise bias, we perform a \commanderthree\ run using half-mission splits with odd-numbered scans and even-numbered scans being analyzed in runs labeled HM1 and HM2. Each of these chains were performed using the same data as in the main \cosmoglobe\ chain, with 200 samples each.\footnote{These products can be found at \url{cosmoglobe.uio.no}.}
The highest quality of similar half-mission splits that are publicly available are from the \Planck\ PR3 analysis, as discussed by \citet{planck2016-l04}. We therefore compute power spectra from the PR3 results and compare them directly with the \cosmoglobe\ HM splits.

\begin{figure}
        \centering
	\includegraphics[width=\linewidth]{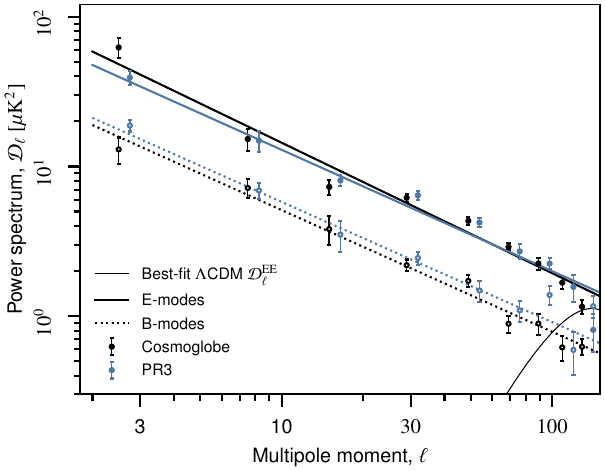}
        \caption{
		Half-mission cross spectra for polarized synchrotron emission as evaluated for \Planck\ PR3 (blue) and \cosmoglobe\ DR1 (black). Filled circles correspond to $E$-modes, while empty circles correspond to $B$-modes. The thick solid and dashed lines are the best-fit $E$- and $B$-mode power law fits to synchrotron, and the thin black line is the $\Lambda$CDM prediction for $E$-modes.
        }
        \label{fig:spectra}
\end{figure}

We perform power spectrum estimation using a so-called pseudo-spectrum code called \texttt{NaMaster} \citep{namaster}, coupled with the \Planck\ 2018\ common polarization mask ($f_\mathrm{sky}=0.78$) and $1^\circ$ apodization.
To quantify the uncertainty, we take the cross spectrum for each pair of Gibbs samples from HM1 and HM2 respectively, and report the 68\,\% confidence intervals on this posterior. The power spectra are displayed in Fig.~\ref{fig:spectra}, and the standard deviation is computed using the within-bin variance of each bin, with the posterior standard deviation of the Gibbs chain added in quadrature for the \cosmoglobe\ spectra.
Other than the very lowest and very highest multipole bins, there is good per-multipole agreement between both the PR3 and \cosmoglobe\ DR1 spectra. For comparison, we plot the $E$-mode power spectrum predicted by the \Planck\ 2018 cosmological parameters.

Following \citet{planck2016-l04}, we perform power law fits to the power spectra of the form ${\mathcal D_\ell^{EE/BB}=A_{\mathrm{s}}^{{EE/BB}}(\ell/80)^\alpha}$, using multipoles $\ell\in[2,140]$. The 68\,\% confidence intervals for each quantity, including the $A_{\mathrm{s}}^{BB}/A_{\mathrm{s}}^{EE}$ ratio, are reported in Table~\ref{tab:synch_powspec}. The primary differences between the fits to the two datasets are $\sim$\,$2\,\sigma$ discrepancies in the $A_{\mathrm{s}}^{BB}$ and $\alpha_s^{EE}$ fits, while all others are consistent within $\simeq0.5~\sigma$. The primary drivers of these differences are lower $\mathcal D_\ell^{BB}$ and higher $\mathcal D_\ell^{EE}$ in the lowest bins.

A question of special interest is the ratio of synchrotron $B$-mode to $E$-mode power, which has been consistently noted to be less than one \citep{page2007,planck2014-a12,planck2016-l04,krachmalnicoff2018,QUIJOTE_IV,eimer2023}. The physical mechanism for this has been discussed in the context of Galactic magnetic fields and polarized thermal dust (e.g., \citealp{kandel2017}, \citealp{so_galsci}, \citealp{vacher2023}), but similar mechanisms are likely to be in play for synchrotron polarization. In Table~\ref{tab:synch_powspec}, we find a value of $A_{\mathrm{s}}^{BB}/A_{\mathrm{s}}^{EE}$ of 0.40 for the \cosmoglobe\ splits and 0.46 for PR3. In this respect, we note that our value of $A_{\mathrm{s}}^{BB}/A_{\mathrm{s}}^{EE}$ for \Planck\ 2018 is higher than that reported by \citet{planck2016-l04} of 0.34, despite using nearly identical methodology. The origin of this discrepancy is not yet understood, but it must necessarily be associated with details in  the power spectrum estimation algorithm. 

\begin{table}
\newdimen\tblskip \tblskip=5pt
	\caption{Best-fit power law parameters to the synchrotron estimates evaluated at 30\,GHz, using half-mission cross-spectra evaluated with \texttt{NaMaster}.}
\label{tab:synch_powspec}
\vskip -5mm
\footnotesize
\setbox\tablebox=\vbox{
 \newdimen\digitwidth
 \setbox0=\hbox{\rm 0}
 \digitwidth=\wd0
 \catcode`*=\active
 \def*{\kern\digitwidth}
  \newdimen\dpwidth
  \setbox0=\hbox{.}
  \dpwidth=\wd0
  \catcode`!=\active
  \def!{\kern\dpwidth}
  \halign{\hbox to 1.8cm{#\leaderfil}\tabskip 2em&
    \hfil$#$\hfil \tabskip 2em&
    \hfil$#$\hfil \tabskip 0em\cr
\noalign{\doubleline}
	\omit\hfil \hfil& \mathrm{PR3} & \cosmoglobe\cr
\noalign{\vskip 3pt\hrule\vskip 5pt}
	$A_{\mathrm{s}}^{EE}$ [$\mathrm{\mu K^2}$]                & \phantom{-}2.39 \pm 0.07 & \phantom{-}2.35 \pm 0.05 \cr
	$A_{\mathrm{s}}^{BB}$ [$\mathrm{\mu K^2}$]                & \phantom{-}1.09 \pm 0.06 & \phantom{-}0.94 \pm 0.04 \cr
	$A_{\mathrm{s}}^{BB}/A_{\mathrm{s}}^{EE}$ & \phantom{-}0.46 \pm 0.03 & \phantom{-}0.40 \pm 0.02 \cr
	$\alpha_s^{EE}$            & -0.81 \pm 0.02 & -0.87 \pm 0.02 \cr
	$\alpha_s^{BB}$            & -0.80 \pm 0.03 & -0.81 \pm 0.03 \cr
\noalign{\vskip 5pt\hrule\vskip 5pt}}}
\endPlancktablewide
\end{table}

\section{Polarized synchrotron spectral indices}
\label{sec:specvar}

Next, we turn our attention to the spectral energy density of polarized synchrotron emission, which we in this section will assume scales as a perfect power law in frequency, $\nu^{\beta_{\mathrm{s}}}$. The determination of $\beta_{\mathrm{s}}$ across the sky has already been studied in detail using several different data combinations and methods. Although the small-scale details vary, nearly every analysis has found $\beta_\mathrm s\simeq-2.8$ in the Galactic plane and $\beta_\mathrm s\simeq-3.3$ in high Galactic latitudes \citep{fuskeland2014,krachmalnicoff2018,fuskeland:2019,weiland:2022}, with the exception of QUIJOTE \citep{QUIJOTE_IV,QUIJOTE_VIII}, who find a slightly flatter spectral index along the Galactic plane.
Both \citet{fuskeland2014} and \citet{weiland:2022} report oscillations with Galactic longitude close to the Galactic plane, but high-latitude regions variations are more difficult to determine, and tend to depend on the specific dataset chosen and the analysis method chosen.


In order to probe the stability of our results with respect to algorithmic details, we perform two different analyses in the following, namely first a so-called $T$--$T$ plot analysis in Sect.~\ref{sec:tt_plot} and then a Gibbs sampling analysis using \commanderone\ \citep{eriksen2008} in Sect.~\ref{sec:comm1}. Section~\ref{sec:tt_plot} focuses on pairs of channels to better isolate potential unmodeled systematic effects, while Sect.~\ref{sec:comm1} uses all \WMAP\ and LFI channels plus \Planck\ 353\,GHz to maximize the joint statistical weight of all available data.

While the \cosmoglobe\ pipeline does produce samples of $\beta_{\mathrm{s}}$ as part of the main Gibbs chain, the full marginal signal-to-noise ratio per pixel with respect to $\beta_{\mathrm{s}}$ is generally very low due to degeneracies with respect to gain and bandpass uncertainties \citep{bp07,bp09}, and this makes it susceptible to systematic uncertainties. When sampled fully unconstrained, $\beta_{\mathrm{s}}$ often drifts to obviously non-physical values at high Galactic latitudes, essentially accounting for low-level gain and bandpass errors, and this in turns permits even the gain to drift into unphysical parameter regions. To avoid this, the default \cosmoglobe\ DR1 processing adopts an informative $\beta_\mathrm s\sim\mathcal N(-3.15, 0.05)$ prior sampling \citep{watts2023_dr1}. This choice has direct implications for the estimates of $\beta_{\mathrm{s}}$ presented in this section, which are derived from the \cosmoglobe\ frequency maps. To at least partially quantify this effect, we consider various different prior choices in Sect.~\ref{sec:comm1}, and we account for shifts resulting from different prior means into our final uncertainties.


\subsection{$T$--$T$ plot analysis}
\label{sec:tt_plot}

\begin{figure}
        \centering
        \includegraphics[width=\linewidth]{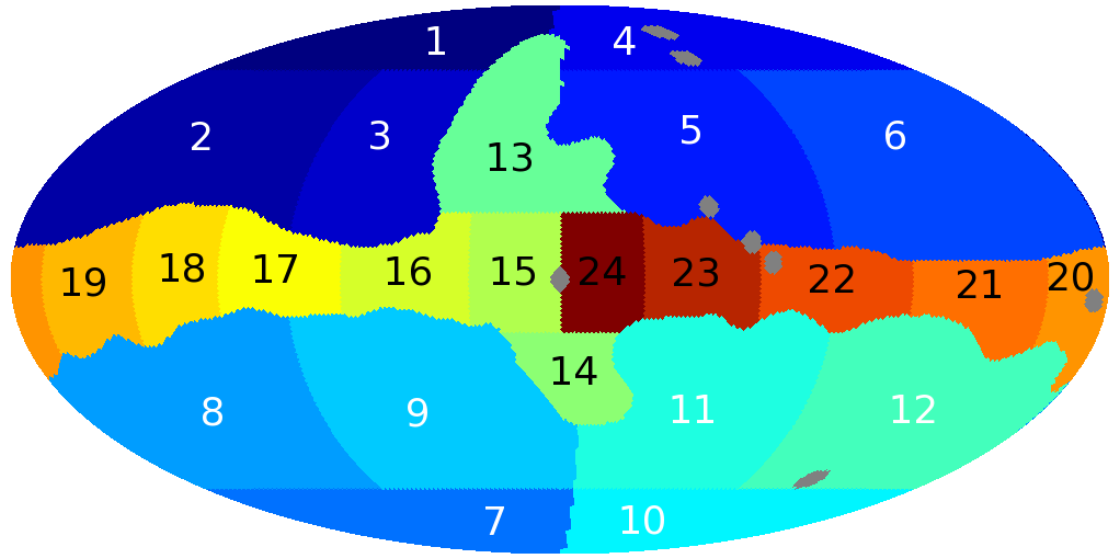}
        \caption{Spectral index regions as defined by \citet{fuskeland2014}. The most prominent point sources are masked out and shown in grey circular areas.
        }
        \label{fig:regions}
\end{figure}

\begin{figure*}
\centering
\includegraphics[width=\linewidth]{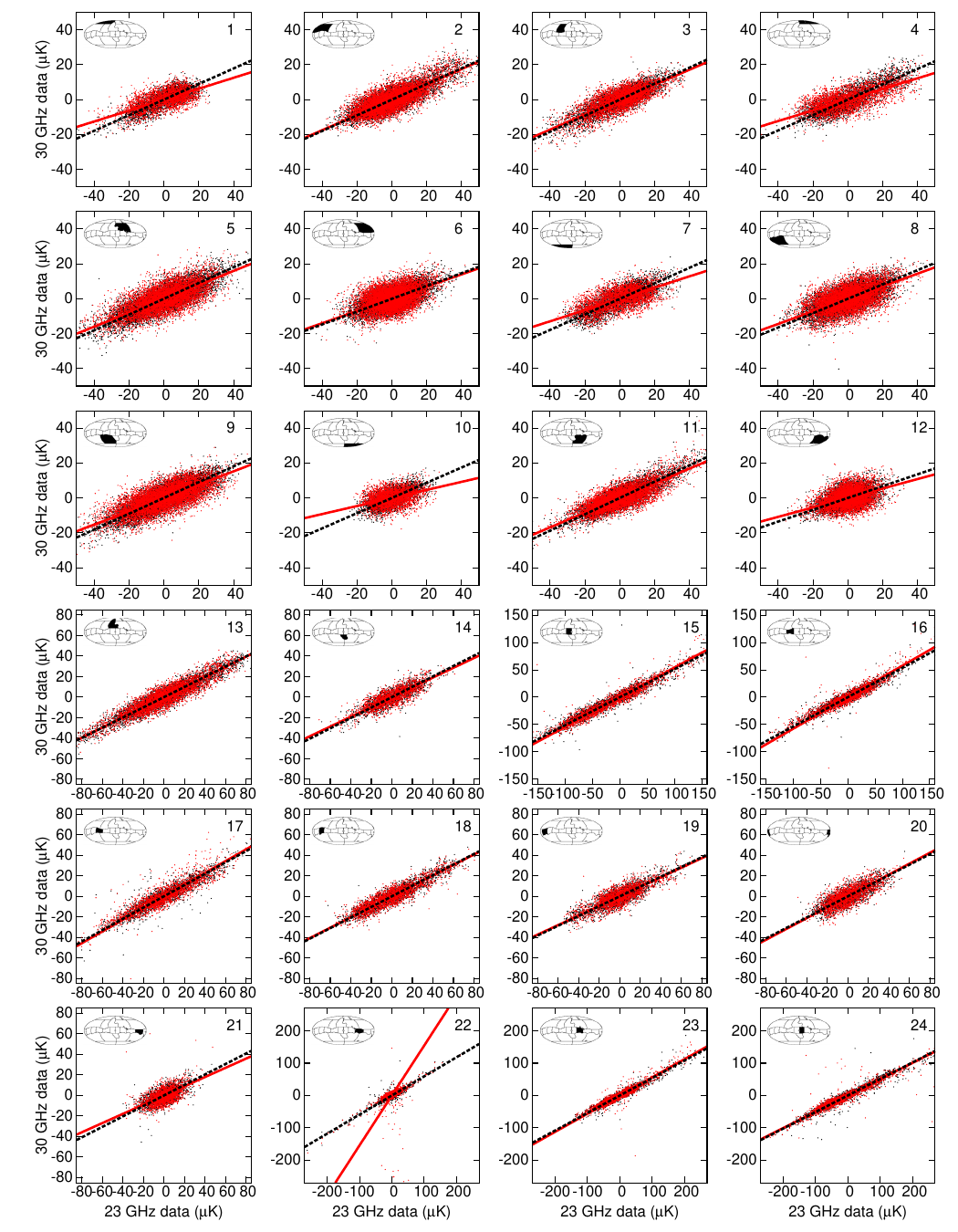}
\captionof{figure}{$T$--$T$ plots for Stokes $Q$ and $U$ maps of the \Cosmoglobe\ DR1 \K-band versus the \Cosmoglobe\ DR1 30\,GHz (black) and the \WMAPnine\ \K-band versus \Planck\ PR3 30\,GHz (red) for all regions. The horizontal (solid and dotted) lines indicate the corresponding inverse variance weighted values of the spectral index, averaged over rotation angle, and in the \Cosmoglobe\ case also weighted over samples.}
\label{fig:cos30_beta_bigscatter}
\end{figure*}

\begin{figure*}
        \centering
        \includegraphics[width=0.99\linewidth]{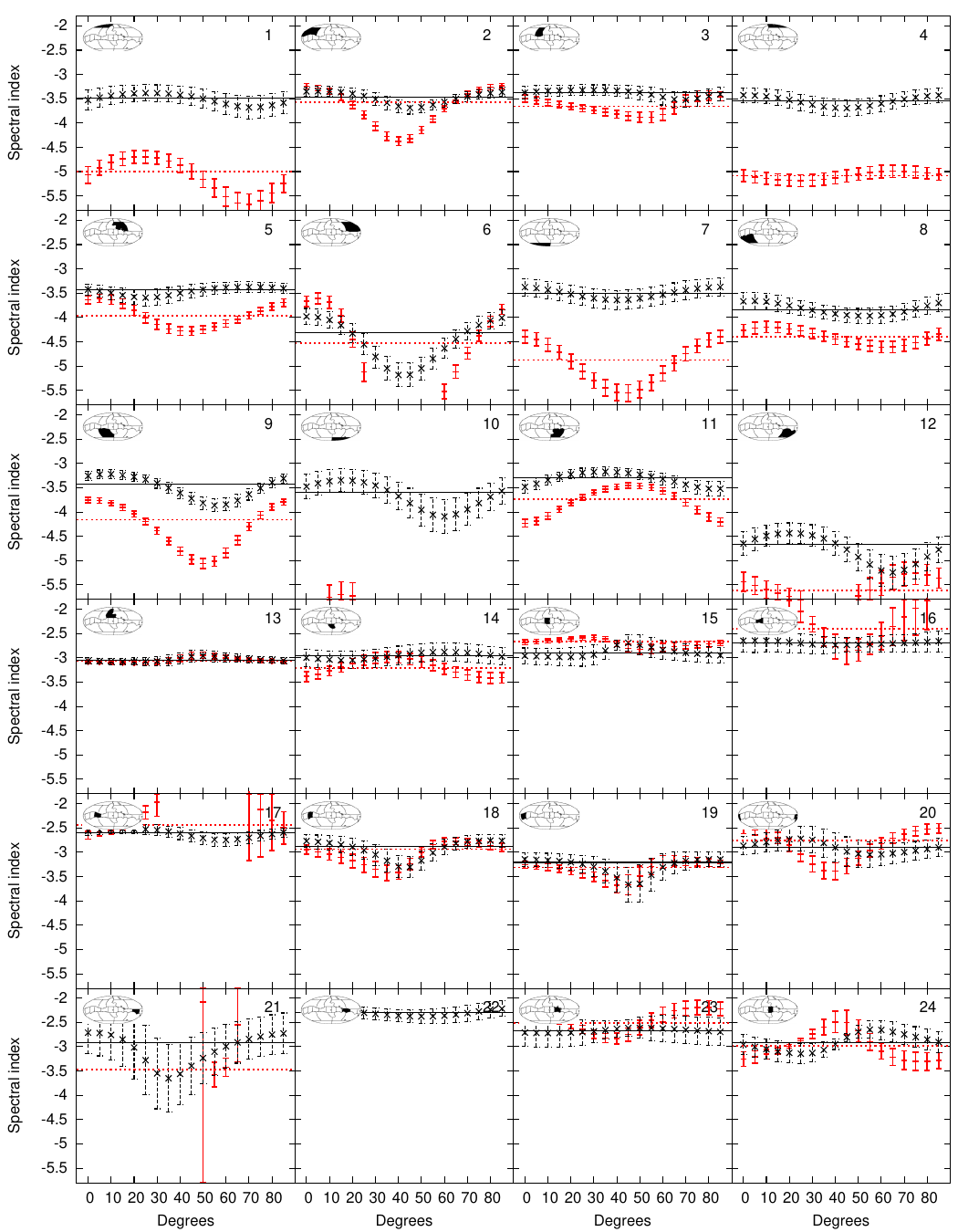}
        \caption{Synchrotron spectral index as a function of rotation angle, computed using $T$--$T$ plot between the \Cosmoglobe\ DR1 \K-band and the \Cosmoglobe\ DR1 30\,GHz (black) compared to the spectral index using the \WMAPnine\ \K-band and \Planck\ PR3 30\,GHz (red) for all regions. The horizontal (solid and dotted) lines indicates the corresponding inverse variance weighted values of the spectral index, averaged over rotation angle, and in the \Cosmoglobe\ case also samples.}
        \label{fig:cos30_beta_bigalpha}
\end{figure*}

\begin{figure}
        \centering
	\includegraphics{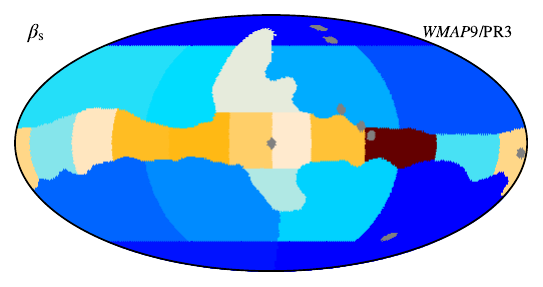}\vspace{-0.25cm}\\
	\includegraphics{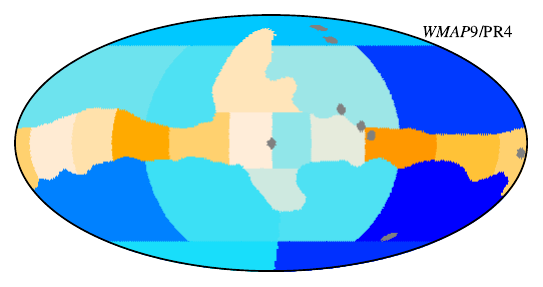}\vspace{-0.25cm}\\
	\includegraphics{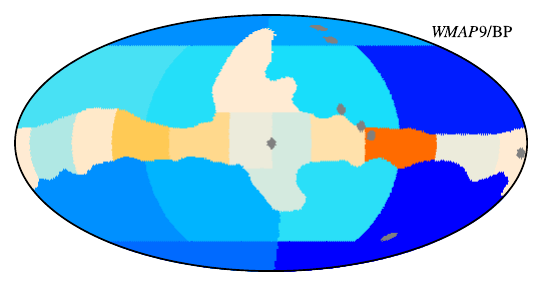}\vspace{-0.25cm}\\
	\includegraphics{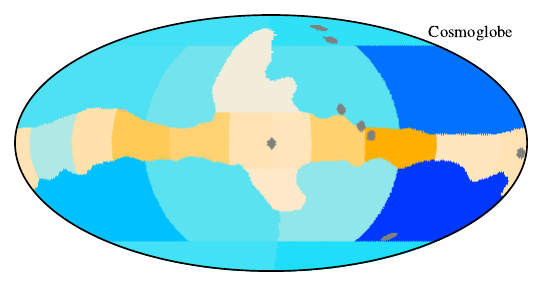}\vspace{-0.25cm}\\
	\hspace{0.25cm}\includegraphics{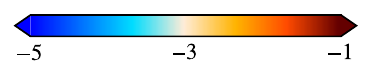}
	\caption{Spatial variation of the synchrotron spectral index, computed using $T$--$T$ plot between the (from top to bottom) \WMAPnine\ \K-band and \Planck\ PR3 30\,GHz, \WMAPnine\ \K-band and \Planck\ PR4 30\,GHz, \WMAPnine\ \K-band and \BP\ 30\,GHz, and \Cosmoglobe\ \K-band and \Cosmoglobe\ 30\,GHz. The spectral index is inverse variance weighted over rotation angle, and in the \Cosmoglobe\ case also samples.}
        \label{fig:TT_beta_maps}
\end{figure}
\begin{figure}
        \centering
        \includegraphics[width=\linewidth]{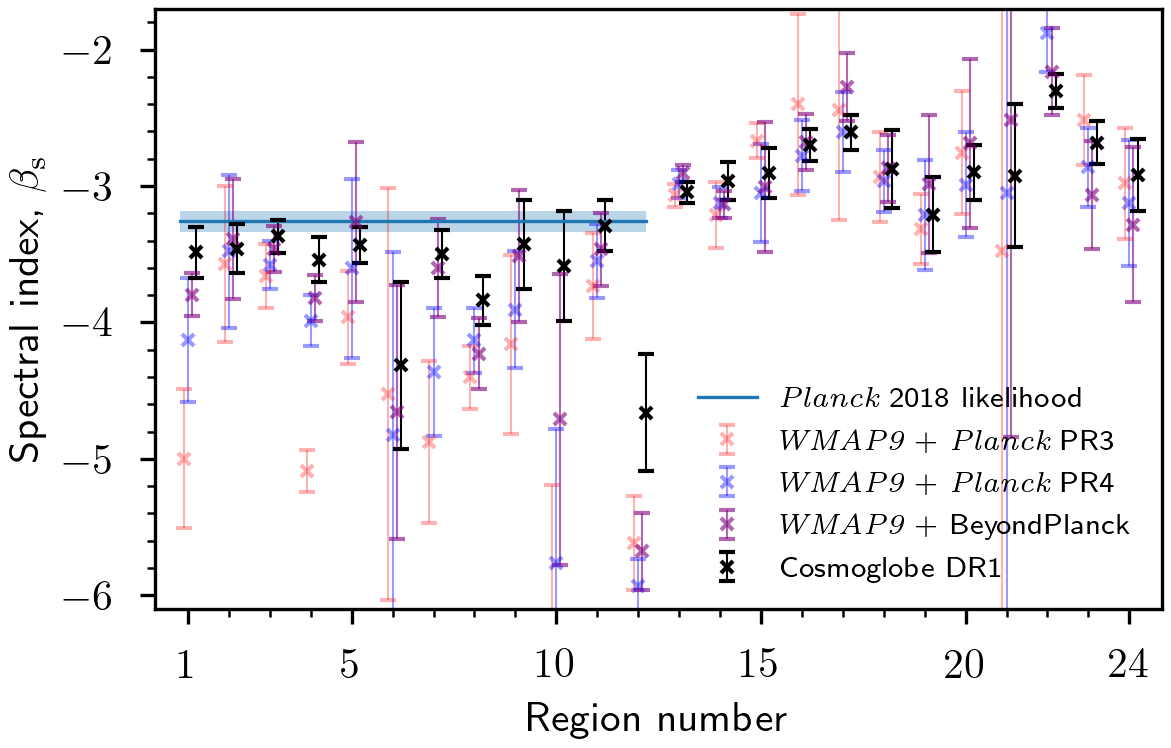}
        \caption{Synchrotron spectral index as a function of region number, computed using $T$--$T$ plot between the \WMAPnine\ \K-band and \Planck\ PR3 30\,GHz (red), \WMAPnine\ \K-band and \Planck\ DR4 30\,GHz (blue), \WMAPnine\ \K-band and \BP\ 30\,GHz (purple), and \Cosmoglobe\ \K-band and \Cosmoglobe\ 30\,GHz (black). The spectral index is inverse variance weighted over rotation angles, and samples. The horizontal line in the high latitude regions corresponds to the estimated spectral index values from the \Planck\ 2018 likelihood analysis \citep{planck2016-l05}. }
        \label{fig:cos30_beta_region}
\end{figure}

The $T$--$T$ plot analysis can be used for pairs of datasets in a model-independent way, allowing for us to probe the difference between different datasets without making strong assumptions on the underlying physical model. This method can also easily be adapted to probe the dependence on polarization angle, and it can be used to study the orientation dependence of $\beta_\mathrm s$ with respect to unmodeled instrumental effects, as discussed by \citet{wehus:2013}.

\begin{figure*}
	\centering
	\includegraphics[width=0.99\textwidth]{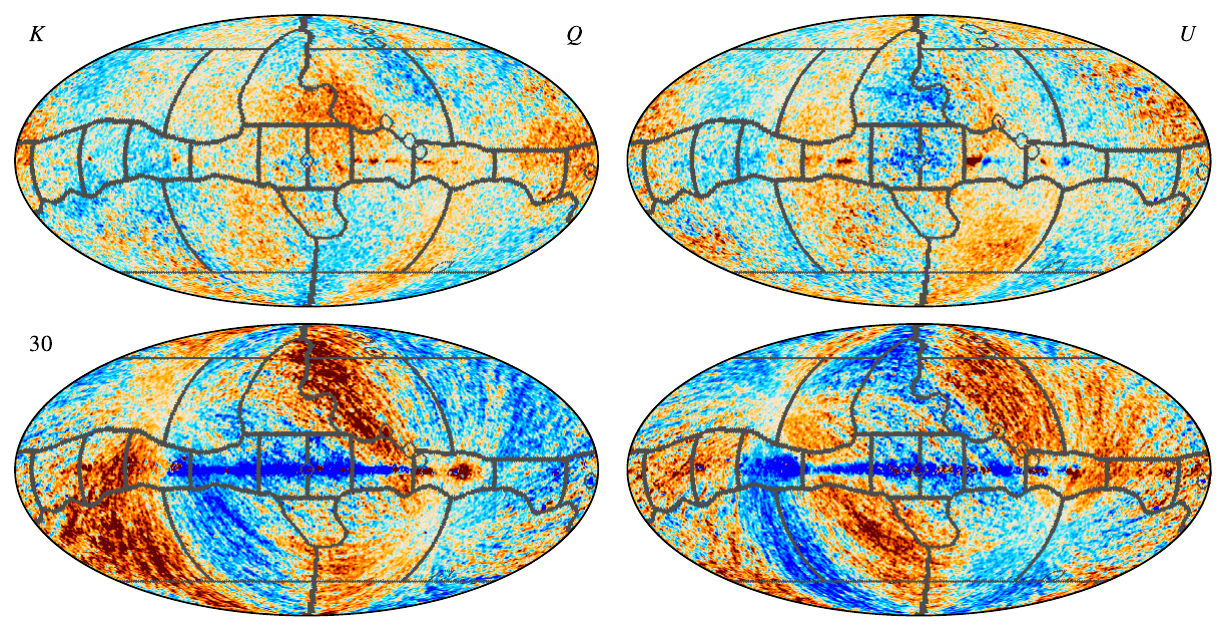}\\
        \includegraphics[width=0.25\textwidth]{figures/cbar_5uK.pdf}
	\caption{Differences between the \cosmoglobe\ and official \WMAP\ 23\,GHz (\emph{top row}) and between \cosmoglobe\ and \Planck\ 2018 30\,GHz (\emph{bottom row}) maps with spectral index regions overlaid. Columns show Stokes $Q$ and $U$ parameters.
		}
       \label{fig:diff_regions}
\end{figure*}

\subsubsection{Method}
\label{sec:tt_plot_method}

Following \citet{fuskeland2014} and \citet{fuskeland:2019}, we apply linear regression via the $T$--$T$ plot method, in which spectral indices can be estimated over extended regions with approximately constant spectral indices. Here we use the regions labeled in Fig.~\ref{fig:regions}. In this approach, our data model reads
\begin{equation}
	\boldsymbol m_\nu = \boldsymbol m_{\nu_0}\left(\frac\nu{\nu_0}\right)^{\beta_\mathrm s}+c_\nu+\boldsymbol n_\nu ,
\end{equation}
where $\boldsymbol m_\nu$ is a spatially varying amplitude map at frequency $\nu$; $\nu_0$ is a reference frequency; $\beta_\mathrm s$ is a power law index; $c_\nu$ the spatially constant offset per band; and $\boldsymbol n_\nu$ represents Gaussian noise. In the case of noiseless data with no offset, the spectral index may be estimated from any two frequency maps, $\nu_1$ and $\nu_2$, using a simple ratio,
\begin{equation}
	\frac{m_{\nu_1,p}}{m_{\nu_2,p}}
	=\left(\frac{\nu_1}{\nu_2}\right)^{\beta_{\mathrm s,p}}
	\Rightarrow
	\beta_{\mathrm s,p}=\frac{\ln(m_{\nu_1,p}/m_{\nu_2,p})}{\ln(\nu_1/\nu_2)}.
\end{equation}
In the case where one map is much noisier than the other, the standard $T$--$T$ plot method involves performing  a linear regression $\boldsymbol m_{\nu_1}=a\boldsymbol m_{\nu_2}+b$, and associates $\beta_\mathrm s$ with $\ln a/\ln(\nu_1/\nu_2)$. More care must be taken when the noise amplitudes in both maps are comparable to each other, so we adopt the effective variance method of \citet{orear1982} as implemented by \citet{fuskeland2014}.

For the $T$--$T$ plot  analysis, we focus exclusively on bands between 23 and 33\,GHz. Following \citet{fuskeland2014}, we use the \WMAP\ \K\ and \Ka\ band Stokes $Q$ and $U$ parameter maps at $23$\,GHz and $33$\,GHz. The respective effective frequencies used are $22.45$\,GHz and $32.64$\,GHz. The maps originally at a \healpix\footnote{\url{http://healpix.sourceforge.net}} pixelization of $N_\textrm{side}=512$ are downgraded to $N_\textrm{side}=64$ and smoothed to a common resolution of $1^\circ$ FWHM.
The \Planck\ data products used are the $30$\,GHz Stokes $Q$ and $U$ maps, with an effective frequency of $28.4$\,GHz. Both the \bp\ and \cosmoglobe\ products are natively at $N_\textrm{side}=512$, while for the \Planck\ PR4, it is 1024.

As described by \citet{fuskeland:2019}, a systematic uncertainty that takes into account the variation of $\beta_\mathrm s$ over rotation angle, $[ \max(\beta_{\mathrm s,\alpha}) - \min(\beta_{\mathrm s,\alpha}) ] /2$, is added in quadrature to the statistical uncertainty.
The uncertainty of the spectral indices is calculated as the minimum of the uncertainties in each rotation angle and region. 
For the \Cosmoglobe\ analyses we have a whole suite of maps represented by the individual samples instead of just one mean map. Here the standard deviation of the spectral indices of the samples is also added in quadrature to represent an additional systematic uncertainty. This enables us to have a better propagation of uncertainties from the maps to the final spectral indices.

\subsubsection{Results}
\label{sec:tt_plot_results}

\begin{figure}
        \centering
        \includegraphics[width=\linewidth]{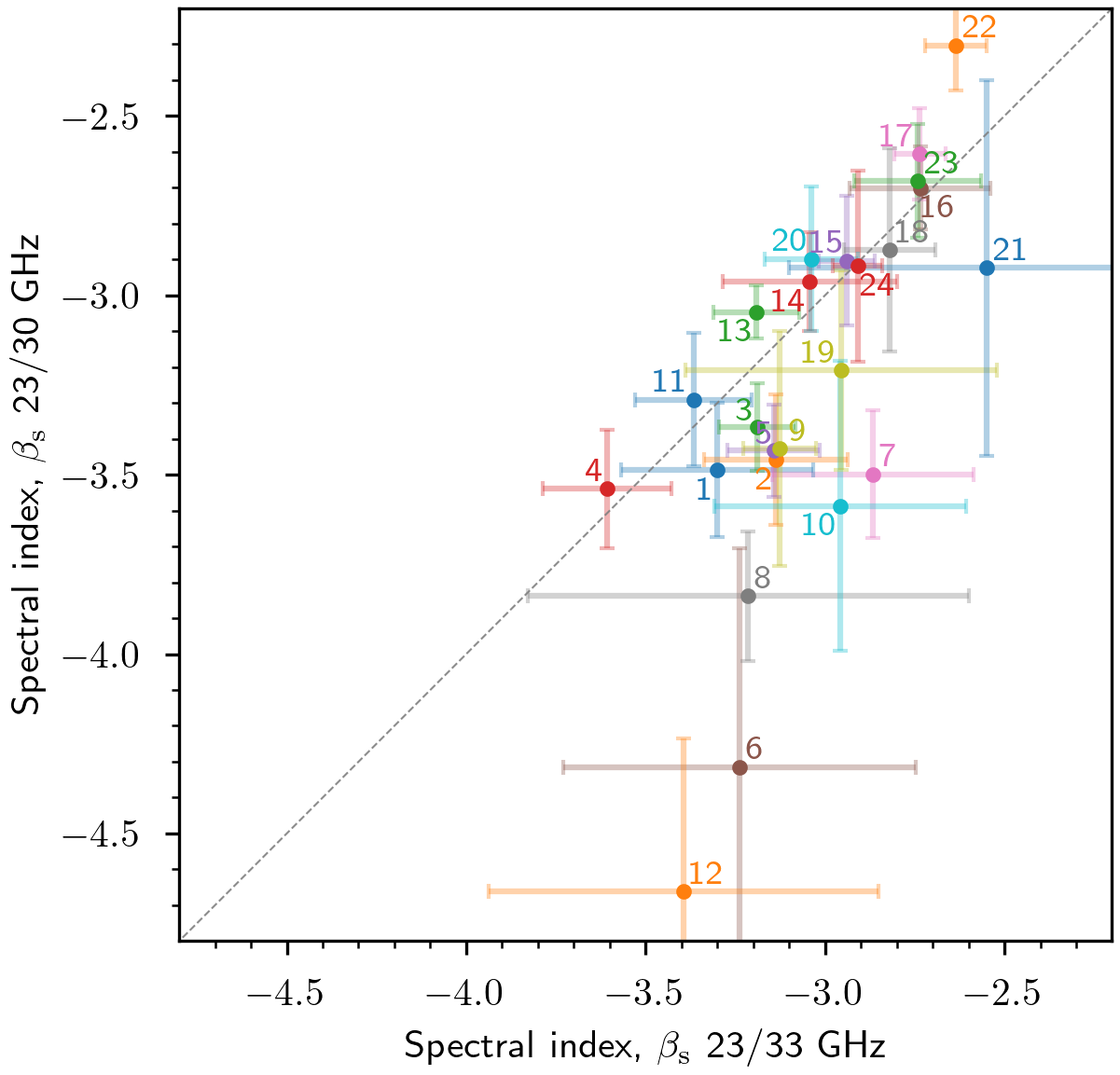}
        \caption{
		Synchrotron spectral index computed using the $T$--$T$ plot method with the \Cosmoglobe\ DR1 \K-band (23\,GHz) and 30\,GHz data versus \Cosmoglobe\ DR1 \K-band and \Ka-band (33\,GHz) data.}
        \label{fig:cos30_xyplot}
\end{figure}

\begin{figure}
	\centering
	\includegraphics{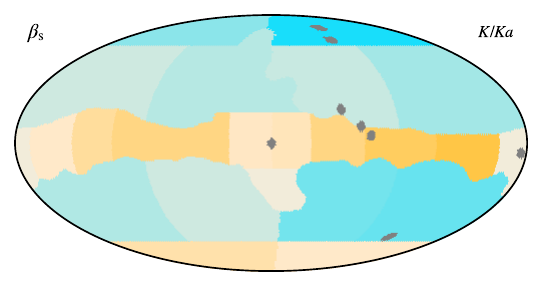}\vspace{-0.25cm}\\
	\hspace{0.25cm}\includegraphics{figures/cbar_beta_wide.pdf}
	\caption{Spatial variation of the synchrotron spectral index, computed using the $T$--$T$ plot method between the \Cosmoglobe\ DR1 \K- and \Ka-band. The spectral index is inverse variance weighted over rotation angle and samples.}
        \label{fig:beta_map}
\end{figure}

\begin{figure}
	\centering
	\includegraphics[width=\columnwidth]{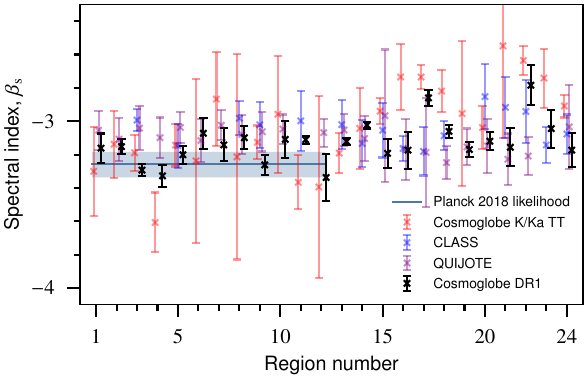}
	\caption{Spectral index from \cosmoglobe\ DR1 spectral index estimation, $T$--$T$ plots with \K/\Ka, the QUIJOTE estimates, and the CLASS estimates.
	}
	\label{fig:beta_comp}
\end{figure}

We start by showing the full set of $T$--$T$ scatter plots between \WMAP\ $K$-band and LFI 30\,GHz in Fig.~\ref{fig:cos30_beta_bigscatter}. Red points show results for the official maps, while black points show the new \cosmoglobe\ maps. Dashed black and solid red lines show best-fit linear fits, respectively. Overall, the two sets of point clouds appear to be in reasonable good agreement at this visual level. Perhaps the most notable features are the fact that while the best-fit lines are generally overlapping in most low Galactic latitude regions, the \cosmoglobe\ slopes are generally a bit shallower than the official maps at high Galactic regions. The most discrepant case is region~22 along the Galactic plane, in which the best-fit line for the official maps are poorly aligned with the diffuse cloud, indicating that a substantial fraction of the pixels have anomalous values. This is typically indicative of strong systematic effects, for instance incomplete temperature-to-polarization leakage.

The information in Fig.~\ref{fig:cos30_beta_bigscatter} is significantly compressed in Fig.~\ref{fig:cos30_beta_bigalpha}, which shows the best-fit spectral index as a function of polarization angle; for full discussion of this type of plots, see, e.g., \citet{wehus:2013} and \citet{fuskeland:2019}. Variations in these curves could in principle be indicative of true spectral variations between the Stokes $Q$ and $U$ parameters, but very large variations are typically associated with uncorrected instrumental systematic errors. Comparing the \cosmoglobe\ (black curves) and official maps (red curves), we generally see that the former show less internal variations than the latter, suggesting greater consistency between the Stokes parameters. For region~22 we see that the official maps fall outside the plotted range. As noted above, we inverse-variance weigh these results when reporting final co-added spectral indices in the following, and we include the Stokes' variation in the final uncertainties.

Figure~\ref{fig:TT_beta_maps} shows final spectral index estimates of this type for four different generations of \WMAP\ \K-band and \Planck\ 30\,GHz data in the form of regionalized sky maps. The top three maps show results derived from the official \WMAP\ \K-band map versus \Planck\ PR3, PR4, and \bp\ 30\,GHz. The bottom map shows the results using the improved maps from the current \cosmoglobe\ analysis. While all cases are inverse variance weighted over rotation angle, the \cosmoglobe\ case is additionally averaged over posterior samples, accounting for low-level instrumental uncertainties. Some of the brightest point sources have been masked out using grey.

We see in this figure an overall narrowing in the range of spectral spectral indices, from regions with extreme values in the top map represented by dark red and blue values, with colors gradually fading in the lower figures. The range in the plot is quite wide for spectral indices, and goes from $-5$ to $-1$. This improvement in the value of the spectral index is especially prominent in the high Galactic regions, as well as region  number 22 along the Galactic plane.  Using the \cosmoglobe\ DR1 data, we find a full-sky inverse variance weighted mean of the spectral index of $\beta_{\mathrm{s}}=-3.07\pm0.07$, while the mean of the high latitude regions 1--14 is $\beta_{\mathrm{s}}=-3.31\pm0.07$.

The same results are also summarized in Fig.~\ref{fig:cos30_beta_region}, with the spectral indices plotted as a function of region number. The different datasets are shown in different colors, and with the same order as the four maps in Fig.~\ref{fig:TT_beta_maps}. Here we have also included one sigma uncertainties as errorbars as described in Sect. \ref{sec:tt_plot_method}. The magnitudes of the error bars are set by the reported white noise level in each dataset except for \cosmoglobe, which also takes into account systematic uncertainties through sample averaging. We also plot the estimated spectral index range for high Galactic regions, in our case represented by regions 1--12, as derived by the \Planck\ 2018 LFI likelihood analysis (see Fig.~22 of \citealp{planck2016-l05}). With each iteration of the data processing, the region estimates draw closer to this result.

When interpreting the variations between different data generations seen in Fig.~\ref{fig:cos30_beta_region}, it is useful to overlay the region boundaries on the raw frequency map differences between the \cosmoglobe\ and official \WMAPnine\ maps \citep{watts2023_dr1}; these are shown in Fig.~\ref{fig:diff_regions} for the \K-band and \Planck\ 30\,GHz maps. As discussed by \citet{watts2023_dr1}, at high Galactic latitudes these large-scale patterns are largely dominated by different gain (for \Planck) and transmission imbalance (for \WMAP) estimates. The most extreme outlier in Fig.~\ref{fig:cos30_beta_region} is region~4, close to the North Galactic Pole, for which the official maps yield a mean spectral index of about $\beta_{\mathrm{s}} \approx -5$. Inspecting Fig.~\ref{fig:diff_regions}, we see that that region has a strong eastward gradient from positive to negative in the \Planck\ 2018 30\,GHz difference, while the opposite is true for \WMAP\ \K-band. Similar considerations hold for many other regions as well. 

Next, we consider similar $T$--$T$ plots for the \K-band and \Ka-band maps. For the \wmap\ data, this analysis was already performed by \citet{fuskeland2014}, so here we only present results using the new \cosmoglobe\ data. First, as a consistency check, in Fig.~\ref{fig:cos30_xyplot} we plot the spectral indices for all 24 regions in the form of \cosmoglobe\ \K/30\,GHz versus \cosmoglobe\ \K/\Ka. We see that for the regions in the north and south Galactic spurs, and along the Galactic plane (regions 13--24), there is a good agreement between the spectral indices obtained by the two pairs of datasets. Regions 1--12, all of which are high-latitude regions with low signal-to-noise, are consistent with the two data combinations. The most prominent outliers, regions 6 and 12, are the regions with the lowest signal-to-noise ratio, and hence most prone to unmodeled effects and noise fluctuations.

In Fig.~\ref{fig:beta_map} the value of the spectral indices is shown for the 24 regions. We see here that the colors are even fainter than the \cosmoglobe\ \K-band versus 30\,GHz values (the bottom figure in Fig.~\ref{fig:TT_beta_maps}), meaning there are fewer outliers with respect to a standard value in the range around $-3$. This is especially visible in the regions with lowest signal to noise, like regions 6, 8 and 12, at high latitude. Taking the inverse variance weighted mean of the spectral indices of all 24 regions, we get $\beta_{\mathrm{s}}=-2.95\pm0.07$, while for the high latitude regions 1--14 we get $-3.20\pm0.10$.



\subsection{Parametric component estimation}
\label{sec:comm1}

\begin{figure}
	\centering
	\includegraphics[width=\columnwidth]{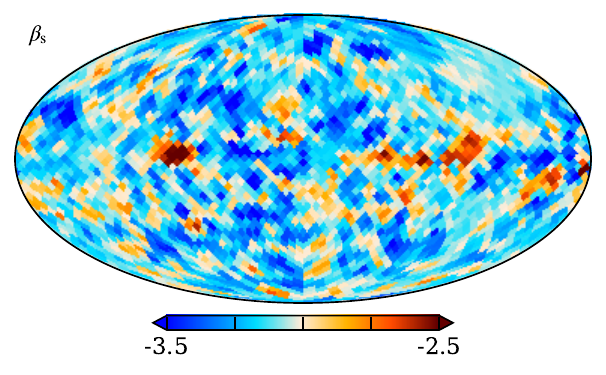}\\
	\includegraphics[width=\columnwidth]{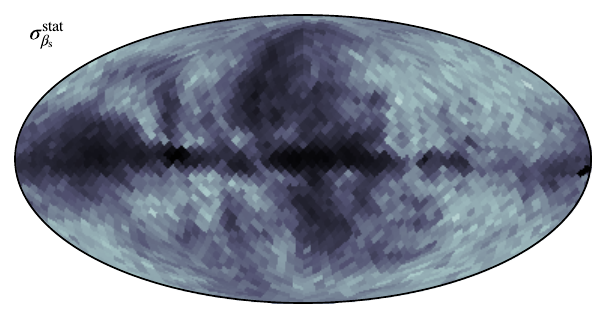}\\
	\includegraphics[width=\columnwidth]{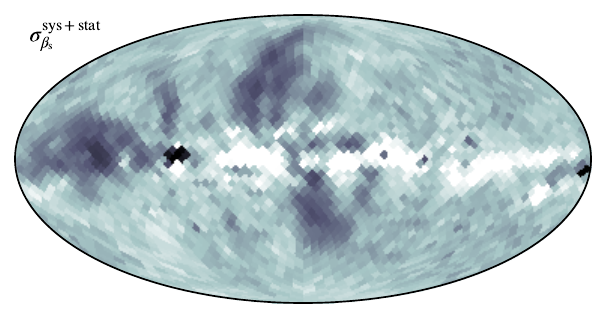}\\
	\includegraphics[width=\columnwidth]{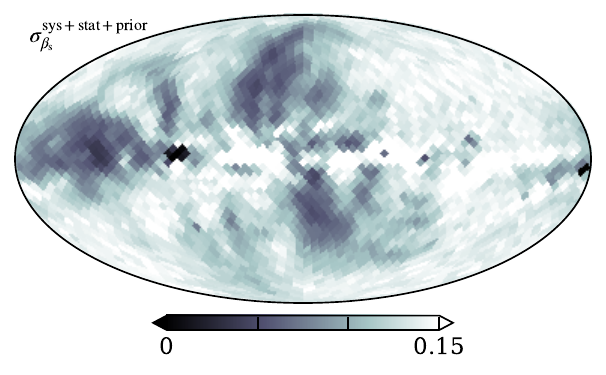}
	\caption{
		\cosmoglobe\ DR1 $\beta_\mathrm s$ constraints. \textit{(First):} Mean spectral index over all samples, \textit{(second):} standard deviation over a single \cosmoglobe\ DR1 sample, \textit{(third):} standard deviation over all \commanderone\ and \cosmoglobe\ DR1 samples, and \textit{(fourth):} difference due to prior choice added in quadrature.}
	\label{fig:beta_16}
\end{figure}

\begin{figure*}
	\begin{center}
	\includegraphics[width=\linewidth]{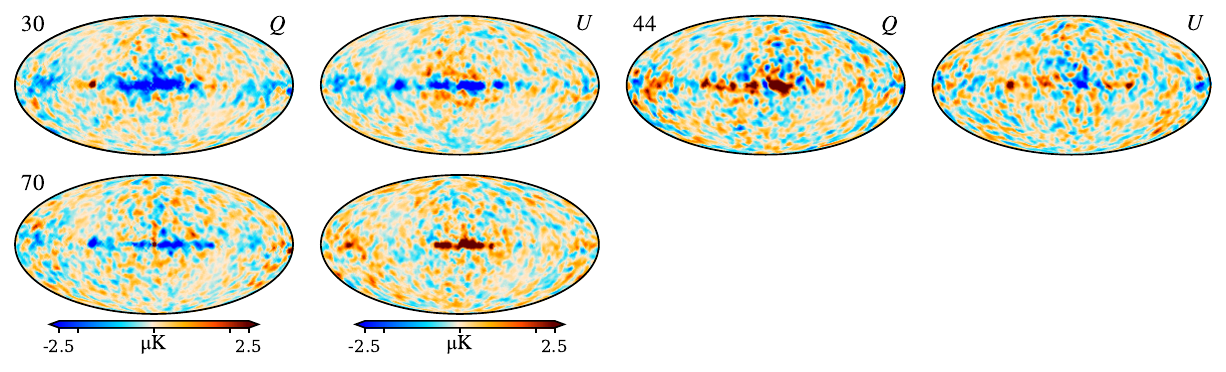}\vspace{-0.3cm}
	\includegraphics[width=\linewidth]{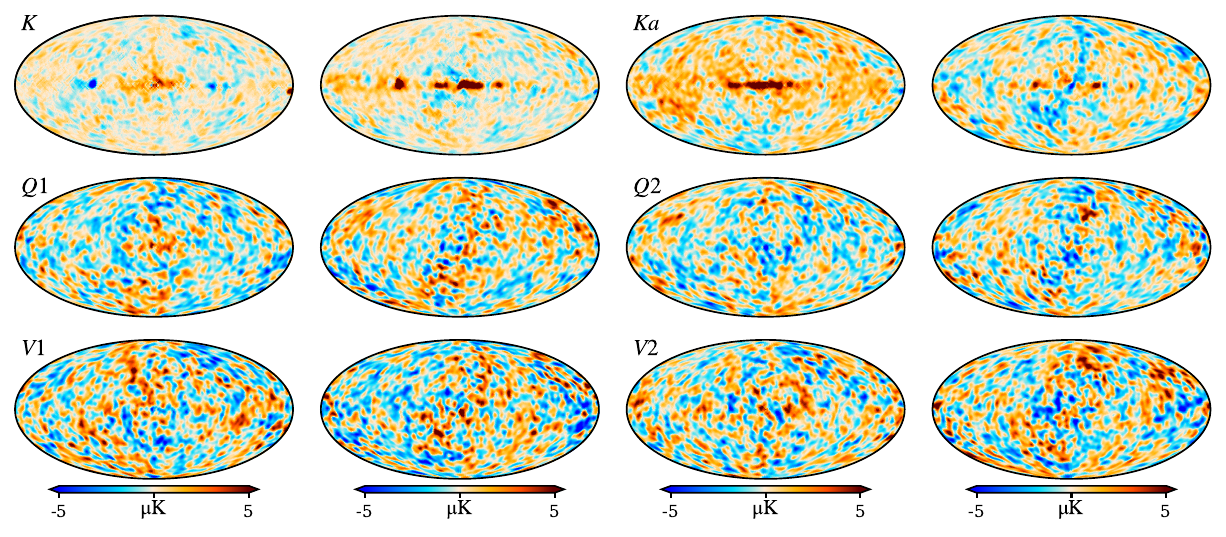}\vspace{-0.3cm}
	\includegraphics[width=\linewidth]{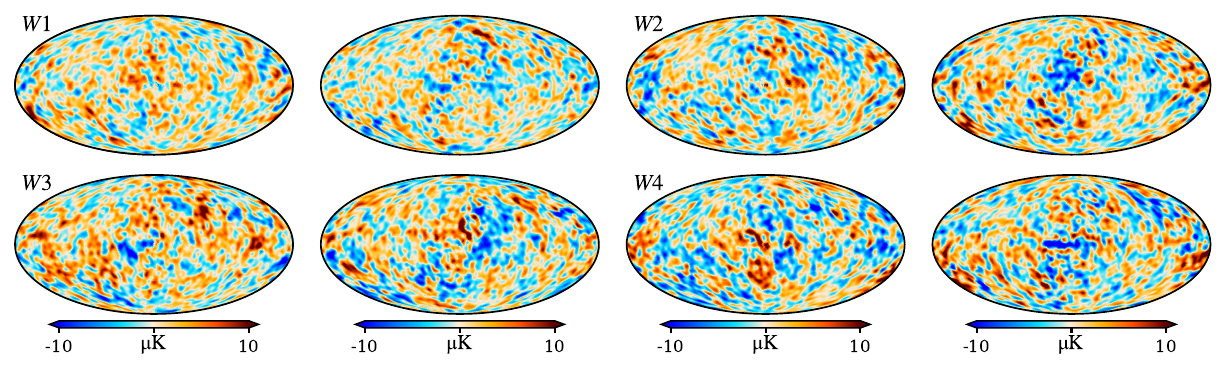}\vspace{-0.3cm}
	\end{center}\vspace{-0.3cm}
	\includegraphics[width=0.5\linewidth]{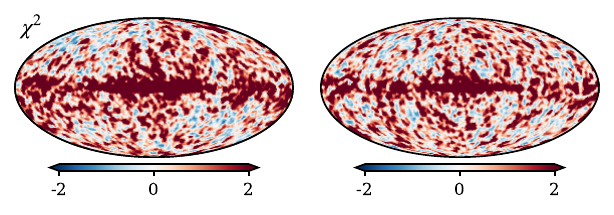}
	\caption{Residual maps and normalized $\chi^2$ in units of $\sigma$ for the \commanderone\ analysis of the \cosmoglobe\ DR1 data. Panels are organized with LFI channels in rows 1--2, \WMAP\ in rows 3--7, and the total reduced $\chi^2$ in row 8.}
	\label{fig:res_QU}
\end{figure*}

In this section we use \commanderone\footnote{\url{https://github.com/Cosmoglobe/Commander1}} to perform pixel-based component separation on maps smoothed to a common resolution of $5^\circ$ and at $N_\mathrm{side}=64$.\footnote{We use \commanderone\ because it is parallelized over pixels and can quickly determine spatial variation for data with common angular resolution; \commanderthree\ is optimized for multi-resolution data, and therefore computationally much more expensive than \commanderone.}
For this analysis, we generate 100 pixel-based \commanderone\ Gibbs samples for each of the 500 \cosmoglobe\ DR1 Gibbs samples produced by \commanderthree. This allows us to decompose the pure statistical error assuming white noise alone for each of the 100 \commanderone\ samples, while changes between each main DR1 sample show the effects of low-level instrumental processing. In these analyses, we use the same polarized bands as in the main \cosmoglobe\ DR1 analysis, namely \WMAP, \Planck\ LFI, and \Planck\ 353\,GHz. We use a prior $\beta_\mathrm s\sim\mathcal N(-3.1, 0.1)$ for the synchrotron spectral index, which is looser than the algorithmic prior used for the main DR1 processing. For thermal dust emission, we adopt the dust temperature map, $T_\mathrm d$, from \citet{planck2014-a12}, coupled with a constant spectral index of $\beta_\mathrm d=1.55$; minor variations in the thermal dust model has only a small impact on the synchrotron spectral index.

Overall, we consider the same data model as in \cosmoglobe\ DR1, but now allow for a spatially varying $\beta_\mathrm s$ in two different forms.
First, we allow $\beta_\mathrm s$ to vary over the same regions as in the $T$--$T$ analysis.
We find that for such large regions, there is only a small effect due to the prior. For instance, when considering very different priors of $\mathcal N(-3.5,0.1)$ and $\mathcal N(-2.7, 0.1)$, we find a maximum difference of $\Delta\beta_{\mathrm{s}}=0.3$ in the lowest signal-to-noise region; generally it is much smaller than this. In the following, we add the resulting differences in quadrature to the uncertainty due to statistical and systematic errors to account for prior ignorance. 

The results from this analysis are summarized in Fig.~\ref{fig:beta_comp} for the nominal priors, and here we also compare our results with corresponding results derived by QUIJOTE \citep{QUIJOTE_VIII} and CLASS \citep{eimer2023}. We choose these data sets because they are to date the most sensitive independent data sets probing a frequency range that is comparable to \WMAP\ and LFI; for comparison S-PASS is much more strongly affected by Faraday rotation \citep{krachmalnicoff2018,fuskeland:2019}. The $\beta_\mathrm s$ maps delivered by the two projects are pixelized at $N_\mathrm{side}=64$ and 32, respectively, with associated uncertainty maps taking into account the expected instrumental noise levels. We take an inverse-weighted average of the respective maps and report the weighted standard error within each region, displayed in Fig.~\ref{fig:beta_comp}. 

In general, the uncertainties in our parametric analysis are smaller than each of the other published results, each for slightly different reasons. First, the $T$--$T$ analyses will inherently have less constraining power than a full likelihood analysis, as this approach only uses two frequency channels via a linear regression, so the uncertainty is determined by the noise level in each frequency channel and the inherent variation within a given sky region. Beyond that, the $T$--$T$ plot in this paper marginalizes over dependence on the polarization angle $\alpha$, and by design accentuates systematic effects, such as beam ellipticity. Finally, the QUIJOTE analysis is most similar in data choice (MFI 11/13\,GHz, \WMAPnine\ \K/\Ka, \Planck\ PR4) and methodology \citep[\texttt{B-SeCRET};][]{b-secret}, but still yields higher uncertainty than the \commanderone\ spectral index region analysis. This is most likely due to different spatial resolution and modeling choices; the \commanderone\ analysis presented here is performed at $5^\circ$ resolution versus the $2^\circ$ resolution per-pixel analysis by \citet{QUIJOTE_VIII}. In addition, \citet{QUIJOTE_VIII} sampled for $\beta_\mathrm d$ and $T_\mathrm{d}$ with priors $\mathcal N(1.55,0.1)$ and $\mathcal N(21,3)$ while using a relatively wide prior on $\beta_\mathrm s$ of $\mathcal N(-3.1, 0.3)$.

At high Galactic latitudes (regions 1--12), there is good agreement between each of the treatments, and all values are consistent with a single constant value. 
In the north and south polar spur (regions 13 and 14) there is excellent agreement between all of the pipelines, while along the Galactic plane (regions 15--24), there are mild discrepancies between the methodologies, in particular with different levels of periodic structure along the Galactic plane.

To more finely probe the spatial variation of $\beta_\mathrm s$, we perform a second \commanderone\ analysis with identical data and model choices, except $\beta_\mathrm s$ is now allowed to vary within each $N_\mathrm{side}=16$ HEALPix pixel. At higher resolutions than this, the spectral index is prior dominated at high Galactic latitudes. In Fig.~\ref{fig:beta_16}, we display the mean of all of the Gibbs samples, along with the standard deviation evaluated per \cosmoglobe\ DR1 Gibbs sample, $\sigma_{\beta_\mathrm s}^\mathrm{stat}$, and the standard deviation over all \cosmoglobe\ DR1 samples and \commanderone\ samples, $\sigma_{\beta_\mathrm s}^\mathrm{sys+stat}$. Thus, $\sigma_{\beta_\mathrm s}^\mathrm{stat}$ is the standard deviation when the input maps themselves are static, and $\sigma_{\beta_\mathrm s}^\mathrm{sys+stat}$ includes variations in the frequency maps themselves, corresponding to underlying instrumental effects, including bandpass, gain, noise characterization, and baseline estimation. The uncertainty due to white noise alone traces the high signal-to-noise regions of the polarized synchrotron, especially the prominent loops and spurs and the Fan region. At high Galactic latitudes, the standard deviation is 0.1, indicating that the posterior uncertainty is limited by the prior. To further quantify this, we have performed additional \commanderone\ runs with priors of $\mathcal N(-3.2,0.1)$ and $\mathcal N(-3.0,0.1)$, and adopt the deviation from the mean in the fiducial analysis as an estimate of the extent of prior domination, $\sigma_{\beta_\mathrm s}^\mathrm{prior}\equiv (\langle \beta_{-3.0}\rangle-\langle\beta_{-3.2}\rangle)/2$. Adding this in quadrature gives the total uncertainty, as shown in the bottom panel of Fig.~\ref{fig:beta_16}.

The uncertainty across the entire Gibbs chain does not merely trace high signal-to-noise regions, and in fact there are variations that exceed the prior surrounding the Galactic center. These variations are primarily due to bandpass and gain uncertainties. Despite the relative increase in noise level when including instrumental effects, the Fan region and the Galactic loops are well-constrained by the data.


As a final quality check, we display the residuals with respect to the \commanderone\ sky model in Fig.~\ref{fig:res_QU}, as well as the total scaled and normalized $\chi^2$. In all bands but \K, 30\,GHz, and \Ka, there are no visible artifacts due to instrumental uncertainty, with each map showing fluctuations consistent with the estimated white noise level calculated in the DR1 processing.  In addition, many of the residuals visible in Fig.~\ref{fig:cg_residuals} have been reduced, demonstrating that polarized synchrotron spectral index variation provides meaningful improvements to the sky model fit.

The most salient remaining residuals are in \K, 30\,GHz, and \Ka. Specifically, \K-band and 30\,GHz are anticorrelated surrounding the Galactic Center, indicating tension between these two high signal-to-noise datasets. This could either be due to genuine mismodeling of the sky, or be due to incompletely modelled instrumental parameters. In particular, the signature is reminiscent of bandpass leakage corrections, which are shown, e.g., in Fig.~9 of \citet{bp09}.

A more persistent residual is found in the Stokes $Q$ \Ka-band map. This feature has appeared in several different analyses, e.g., Fig.~4 of \citet{bp14} and Fig.~8 of \citet{weiland:2022}, but was not as clear without full removal of the poorly measured modes in the final map. The lack of this feature in the corresponding Stokes $U$ map suggests that the effect is not a true Galactic effect, and is in some way due to instrumental processing, or unmodeled systematics; potential sources of this signal are discussed below.

\section{Discussion and conclusion}
\label{sec:conclusion}

We have presented a new state-of-the-art model of polarized synchrotron using the \cosmoglobe\ DR1 data products, for the first time combining all \Planck\ LFI and \WMAP\ frequency maps and thereby leveraging the full statistical power of these data. The polarized synchrotron map as delivered by \citet{watts2023_dr1} has an effective white noise level of $3.4\,\mathrm{\mu K}$, a 29\,\% improvement over the white noise levels of the \WMAP\ \K-band and LFI 30\,GHz maps. This model characterizes the polarized sky signal to within $5\,\mathrm{\mu K}$ at all bands. We have also reproduced the previously-reported $B$-to-$E$-mode ratio in the polarized synchrotron power spectrum. Furthermore, we have shown that physically reasonable spectral indices can be recovered using a variety of methods, which are consistent with both the methods presented in this paper and with previously published results from CLASS and QUIJOTE, confirming variation in $\beta_\mathrm s$ along the Galactic plane, and steepening at high Galactic latitudes.

Our improved processing of the \WMAP\ and LFI data in conjunction with improved polarized synchrotron modeling has allowed us to dig deeper into the underlying differences between the two datasets. We have shown agreement between the $\beta_\mathrm s$ values derived using the $T$--$T$ plot using \K/30\,GHz and \K/\Ka\ data combinations, further evidence that these datasets are now consistent with each other. The remaining unexplained differences between the sky maps and our derived model are primarily in the Galactic center, and have morphologies consistent with bandpass errors and SED complexities.

The least well-understood residual is in the \Ka-band, specifically Stokes $Q$. Of all the known instrumental effects in both \WMAP\ and \Planck\ LFI, the only one that morphologically resembles this residual is the bandpass correction in \K-band, which was previously presented in Fig.~12 of \citet{watts2023_dr1}. The bandpass correction term depends solely on laboratory measurements, presented by \citet{jarosik2003:MAP}. The high statistical weight of \K-band for determining polarized synchrotron could easily lead to an undersubtraction in the \Ka-band maps. Conversely, the nominal bandpass correction to \Ka-band itself is negligible, as all of the \Ka-band radiometers were reported to have nearly identical bandpasses. A full accounting of this effect, including sampling of bandpass differences in the \WMAP\ TODs (not performed by \citealt{watts2023_dr1}) will be performed in future work.

The estimation of $\beta_\mathrm s$ is difficult precisely because the quantity is dependent on differences between different frequencies, which can often exacerbate systematic effects and processing choices. The differences between a sky model and datasets is best determined in the timestreams -- much of the improvement demonstrated in this analysis would not have been possible through a purely map-based analysis. Beyond pure improvement of data quality, this approach allowed for a natural end-to-end propagation of errors, and demonstrated that some of the brightest regions of the sky in fact do not have well-determined spectral indices, despite their high signal-to-instrumental noise ratio. As increasingly stringent uncertainty constraints are becoming necessary in order to measure a non-zero tensor-to-scalar ratio, the use of joint information between experiments with complementary observation strategies will increasingly become a necessity. Future joint analyses, including for example QUIJOTE and CLASS data, will continue to improve our knowledge of the polarized sky.

\input{CG_acknowledgments.tex}

\bibliographystyle{aa}

\bibliography{Planck_bib,CG_bibliography}






\end{document}

%% file: Planck.tex
\def\setsymbol#1#2{\expandafter\def\csname #1\endcsname{#2}}
\def\getsymbol#1{\csname #1\endcsname}

\def\Planck{\textit{Planck}}

\newbox\tablebox    \newdimen\tablewidth
\def\leaderfil{\leaders\hbox to 5pt{\hss.\hss}\hfil}
\def\endPlancktablewide{\tablewidth=\textwidth 
    $$\hss\copy\tablebox\hss$$
    \vskip-\lastskip\vskip -2pt}
\def\tablenote#1 #2\par{\begingroup \parindent=0.8em
    \abovedisplayshortskip=0pt\belowdisplayshortskip=0pt
    \noindent
    $$\hss\vbox{\hsize\tablewidth \hangindent=\parindent \hangafter=1 \noindent
    \hbox to \parindent{$^#1$\hss}\strut#2\strut\par}\hss$$
    \endgroup}
\def\doubleline{\vskip 3pt\hrule \vskip 1.5pt \hrule \vskip 5pt}

\def\L2{\ifmmode L_2\else $L_2$\fi}

\def\DeltaT{\ifmmode \Delta T\else $\Delta T$\fi}
\def\deltat{\ifmmode \Delta t\else $\Delta t$\fi}
\def\fknee{\ifmmode f_{\rm knee}\else $f_{\rm knee}$\fi}
\def\Fmax{\ifmmode F_{\rm max}\else $F_{\rm max}$\fi}
\def\solar{\ifmmode{\rm M}_{\mathord\odot}\else${\rm M}_{\mathord\odot}$\fi}
\def\Msolar{\ifmmode{\rm M}_{\mathord\odot}\else${\rm M}_{\mathord\odot}$\fi}
\def\Lsolar{\ifmmode{\rm L}_{\mathord\odot}\else${\rm L}_{\mathord\odot}$\fi}
\def\inv{\ifmmode^{-1}\else$^{-1}$\fi}
\def\mo{\ifmmode^{-1}\else$^{-1}$\fi}
\def\sup#1{\ifmmode ^{\rm #1}\else $^{\rm #1}$\fi}
\def\expo#1{\ifmmode \times 10^{#1}\else $\times 10^{#1}$\fi}
\def\,{\thinspace}
\def\lsim{\mathrel{\raise .4ex\hbox{\rlap{$<$}\lower 1.2ex\hbox{$\sim$}}}}
\def\gsim{\mathrel{\raise .4ex\hbox{\rlap{$>$}\lower 1.2ex\hbox{$\sim$}}}}

\def\simprop{\mathrel{\raise .4ex\hbox{\rlap{$\propto$}\lower 1.2ex\hbox{$\sim$}}}}
\def\deg{\ifmmode^\circ\else$^\circ$\fi}
\def\pdeg{\ifmmode $\setbox0=\hbox{$^{\circ}$}\rlap{\hskip.11\wd0 .}$^{\circ}
          \else \setbox0=\hbox{$^{\circ}$}\rlap{\hskip.11\wd0 .}$^{\circ}$\fi}
\def\arcs{\ifmmode {^{\scriptstyle\prime\prime}}
          \else $^{\scriptstyle\prime\prime}$\fi}
\def\arcm{\ifmmode {^{\scriptstyle\prime}}
          \else $^{\scriptstyle\prime}$\fi}
\newdimen\sa  \newdimen\sb
\def\parcs{\sa=.07em \sb=.03em
     \ifmmode \hbox{\rlap{.}}^{\scriptstyle\prime\kern -\sb\prime}\hbox{\kern -\sa}
     \else \rlap{.}$^{\scriptstyle\prime\kern -\sb\prime}$\kern -\sa\fi}
\def\parcm{\sa=.08em \sb=.03em
     \ifmmode \hbox{\rlap{.}\kern\sa}^{\scriptstyle\prime}\hbox{\kern-\sb}
     \else \rlap{.}\kern\sa$^{\scriptstyle\prime}$\kern-\sb\fi}
\def\ra[#1 #2 #3.#4]{#1\sup{h}#2\sup{m}#3\sup{s}\llap.#4}
\def\dec[#1 #2 #3.#4]{#1\deg#2\arcm#3\arcs\llap.#4}
\def\deco[#1 #2 #3]{#1\deg#2\arcm#3\arcs}
\def\rra[#1 #2]{#1\sup{h}#2\sup{m}}

\def\dots{\relax\ifmmode \ldots\else $\ldots$\fi}
\def\WHzsr{\ifmmode $W\,Hz\mo\,sr\mo$\else W\,Hz\mo\,sr\mo\fi}
\def\mHz{\ifmmode $\,mHz$\else \,mHz\fi}
\def\GHz{\ifmmode $\,GHz$\else \,GHz\fi}
\def\mKs{\ifmmode $\,mK\,s$^{1/2}\else \,mK\,s$^{1/2}$\fi}
\def\muKs{\ifmmode \,\mu$K\,s$^{1/2}\else \,$\mu$K\,s$^{1/2}$\fi}
\def\muKRJs{\ifmmode \,\mu$K$_{\rm RJ}$\,s$^{1/2}\else \,$\mu$K$_{\rm RJ}$\,s$^{1/2}$\fi}
\def\muKHz{\ifmmode \,\mu$K\,Hz$^{-1/2}\else \,$\mu$K\,Hz$^{-1/2}$\fi}
\def\MJysr{\ifmmode \,$MJy\,sr\mo$\else \,MJy\,sr\mo\fi}
\def\MJysrmK{\ifmmode \,$MJy\,sr\mo$\,mK$_{\rm CMB}\mo\else \,MJy\,sr\mo\,mK$_{\rm CMB}\mo$\fi}
\def\microns{\ifmmode \,\mu$m$\else \,$\mu$m\fi}

\def\muK{\ifmmode \,\mu$K$\else \,$\mu$\hbox{K}\fi}
\def\microK{\ifmmode \,\mu$K$\else \,$\mu$\hbox{K}\fi}
\def\muW{\ifmmode \,\mu$W$\else \,$\mu$\hbox{W}\fi}
\def\kms{\ifmmode $\,km\,s$^{-1}\else \,km\,s$^{-1}$\fi}
\def\kmsMpc{\ifmmode $\,\kms\,Mpc\mo$\else \,\kms\,Mpc\mo\fi}
\providecommand{\sorthelp}[1]{}

%% file: authors.tex
\author{\small
D.~J.~Watts\inst{\ref{oslo}}\thanks{Corresponding author: D.~Watts; \url{duncanwa@astro.uio.no}}
\and
U.~Fuskeland\inst{\ref{oslo}}
\and
R.~Aurlien\inst{\ref{oslo}}
\and
A.~Basyrov\inst{\ref{oslo}}
\and
L.~A.~Bianchi\inst{\ref{milano}}
\and
M.~Brilenkov\inst{\ref{oslo}}
\and
H.~K.~Eriksen\inst{\ref{oslo}}
\and
K.~S.~F.~Fornazier\inst{\ref{saopaulo}}
\and\\
M.~Galloway\inst{\ref{oslo}}
\and
E.~Gjerløw\inst{\ref{oslo}}
\and
B.~Hensley\inst{\ref{jpl}}
\and
L.~T.~Hergt\inst{\ref{ubc}}
\and
D.~Herman\inst{\ref{oslo}}
\and
H.~Ihle\inst{\ref{oslo}}
\and
K.~Lee\inst{\ref{oslo}}
\and
J.~G.~S.~Lunde\inst{\ref{oslo}}
\and
S.~K.~Nerval\inst{\ref{dunlap1},\ref{dunlap2}}
\and\\
M.~San\inst{\ref{oslo}}
\and
N.~O.~Stutzer\inst{\ref{oslo}}
\and
H.~Thommesen\inst{\ref{oslo}}
\and
I.~K.~Wehus\inst{\ref{oslo}}
}
\institute{\small
Institute of Theoretical Astrophysics, University of Oslo, Blindern, Oslo, Norway\label{oslo}
\and
Dipartimento di Fisica, Universit\`{a} degli Studi di Milano, Via Celoria, 16, Milano, Italy\label{milano}
\and
Instituto de Física, Universidade de São Paulo - C.P. 66318, CEP: 05315-970, São Paulo, Brazil\label{saopaulo}
\and
Jet Propulsion Laboratory, California Institute of Technology, 4800 Oak Grove Drive, Pasadena, California, U.S.A.\label{jpl}
\and
Department of Physics and Astronomy, University of British Columbia, 6224 Agricultural Road, Vancouver BC, V6T1Z1, Canada\label{ubc}
\and
David A. Dunlap Department of Astronomy \& Astrophysics, University of Toronto, 50 St. George Street, Toronto, ON M5S 3H4, Canada\label{dunlap1}
\and
Dunlap Institute for Astronomy \& Astrophysics, University of Toronto, 50 St. George Street, Toronto, ON M5S 3H4, Canada\label{dunlap2}
}

%% file: CG_acknowledgments.tex
\begin{acknowledgements}
  We thank the entire \Planck\ and \WMAP\ teams for invaluable support
  and discussions, and for their dedicated efforts through several
  decades without which this work would not be possible. The current
  work has received funding from the European Union’s Horizon 2020
  research and innovation programme under grant agreement numbers
  819478 (ERC; \textsc{Cosmoglobe}), 772253 (ERC;
  \textsc{bits2cosmology}), and 101007633 (Marie Skłodowska-Curie,
  \textsc{CMB-INFLATE}). In addition, the collaboration acknowledges
  support from RCN (Norway; grant no.\ 274990). The research was
  carried out in part at the Jet Propulsion Laboratory, California
  Institute of Technology, under a contract with the National
  Aeronautics and Space Administration (80NM0018D0004).  We
  acknowledge the use of the Legacy Archive for Microwave Background
  Data Analysis (LAMBDA), part of the High Energy Astrophysics Science
  Archive Center (HEASARC). HEASARC/LAMBDA is a service of the
  Astrophysics Science Division at the NASA Goddard Space Flight
  Center.  Some of the results in this paper have been derived using
  the \texttt{healpy} and
  \texttt{HEALPix}\footnote{\url{http://healpix.sf.net}} packages
  \citep{gorski2005, Zonca2019}.  This work made use of
  Astropy:\footnote{\url{http://www.astropy.org}} a
  community-developed core Python package and an ecosystem of tools
  and resources for astronomy \citep{astropy:2013, astropy:2018,
    astropy:2022}.
\end{acknowledgements}

%% file: main.bbl
\begin{thebibliography}{58}
\expandafter\ifx\csname natexlab\endcsname\relax\def\natexlab#1{#1}\fi

\bibitem[{{Abazajian} {et~al.}(2019){Abazajian}, {Addison}, {Adshead}, {Ahmed},
  {Allen}, {Alonso}, {Alvarez}, {Anderson}, {Arnold}, {Baccigalupi}, {Bailey},
  {Barkats}, {Barron}, {Barry}, {Bartlett}, {Basu Thakur}, {Battaglia},
  {Baxter}, {Bean}, {Bebek}, {Bender}, {Benson}, {Berger}, {Bhimani},
  {Bischoff}, {Bleem}, {Bocquet}, {Boddy}, {Bonato}, {Bond}, {Borrill},
  {Bouchet}, {Brown}, {Bryan}, {Burkhart}, {Buza}, {Byrum}, {Calabrese},
  {Calafut}, {Caldwell}, {Carlstrom}, {Carron}, {Cecil}, {Challinor}, {Chang},
  {Chinone}, {Cho}, {Cooray}, {Crawford}, {Crites}, {Cukierman}, {Cyr-Racine},
  {de Haan}, {de Zotti}, {Delabrouille}, {Demarteau}, {Devlin}, {Di Valentino},
  {Dobbs}, {Duff}, {Duivenvoorden}, {Dvorkin}, {Edwards}, {Eimer}, {Errard},
  {Essinger-Hileman}, {Fabbian}, {Feng}, {Ferraro}, {Filippini}, {Flauger},
  {Flaugher}, {Fraisse}, {Frolov}, {Galitzki}, {Galli}, {Ganga}, {Gerbino},
  {Gilchriese}, {Gluscevic}, {Green}, {Grin}, {Grohs}, {Gualtieri}, {Guarino},
  {Gudmundsson}, {Habib}, {Haller}, {Halpern}, {Halverson}, {Hanany},
  {Harrington}, {Hasegawa}, {Hasselfield}, {Hazumi}, {Heitmann}, {Henderson},
  {Henning}, {Hill}, {Hlozek}, {Holder}, {Holzapfel}, {Hubmayr},
  {Huffenberger}, {Huffer}, {Hui}, {Irwin}, {Johnson}, {Johnstone}, {Jones},
  {Karkare}, {Katayama}, {Kerby}, {Kernovsky}, {Keskitalo}, {Kisner}, {Knox},
  {Kosowsky}, {Kovac}, {Kovetz}, {Kuhlmann}, {Kuo}, {Kurita}, {Kusaka},
  {Lahteenmaki}, {Lawrence}, {Lee}, {Lewis}, {Li}, {Linder}, {Loverde},
  {Lowitz}, {Madhavacheril}, {Mantz}, {Matsuda}, {Mauskopf}, {McMahon},
  {McQuinn}, {Meerburg}, {Melin}, {Meyers}, {Millea}, {Mohr}, {Moncelsi},
  {Mroczkowski}, {Mukherjee}, {M{\"u}nchmeyer}, {Nagai}, {Nagy}, {Namikawa},
  {Nati}, {Natoli}, {Negrello}, {Newburgh}, {Niemack}, {Nishino}, {Nordby},
  {Novosad}, {O'Connor}, {Obied}, {Padin}, {Pandey}, {Partridge}, {Pierpaoli},
  {Pogosian}, {Pryke}, {Puglisi}, {Racine}, {Raghunathan}, {Rahlin},
  {Rajagopalan}, {Raveri}, {Reichanadter}, {Reichardt}, {Remazeilles}, {Rocha},
  {Roe}, {Roy}, {Ruhl}, {Salatino}, {Saliwanchik}, {Schaan}, {Schillaci},
  {Schmittfull}, {Scott}, {Sehgal}, {Shandera}, {Sheehy}, {Sherwin},
  {Shirokoff}, {Simon}, {Slosar}, {Somerville}, {Spergel}, {Staggs}, {Stark},
  {Stompor}, {Story}, {Stoughton}, {Suzuki}, {Tajima}, {Teply}, {Thompson},
  {Timbie}, {Tomasi}, {Treu}, {Tristram}, {Tucker}, {Umilt{\`a}}, {van
  Engelen}, {Vieira}, {Vieregg}, {Vogelsberger}, {Wang}, {Watson}, {White},
  {Whitehorn}, {Wollack}, {Kimmy Wu}, {Xu}, {Yasini}, {Yeck}, {Yoon}, {Young},
  \& {Zonca}}]{cmbs4}
{Abazajian}, K., {Addison}, G., {Adshead}, P., {et~al.} 2019, arXiv e-prints,
  arXiv:1907.04473

\bibitem[{{Ade} {et~al.}(2019){Ade}, {Aguirre}, {Ahmed}, {Aiola}, {Ali},
  {Alonso}, {Alvarez}, {Arnold}, {Ashton}, {Austermann}, {Awan}, {Baccigalupi},
  {Baildon}, {Barron}, {Battaglia}, {Battye}, {Baxter}, {Bazarko}, {Beall},
  {Bean}, {Beck}, {Beckman}, {Beringue}, {Bianchini}, {Boada}, {Boettger},
  {Bond}, {Borrill}, {Brown}, {Bruno}, {Bryan}, {Calabrese}, {Calafut},
  {Calisse}, {Carron}, {Challinor}, {Chesmore}, {Chinone}, {Chluba}, {Cho},
  {Choi}, {Coppi}, {Cothard}, {Coughlin}, {Crichton}, {Crowley}, {Crowley},
  {Cukierman}, {D'Ewart}, {D{\"u}nner}, {de Haan}, {Devlin}, {Dicker},
  {Didier}, {Dobbs}, {Dober}, {Duell}, {Duff}, {Duivenvoorden}, {Dunkley},
  {Dusatko}, {Errard}, {Fabbian}, {Feeney}, {Ferraro}, {Flux{\`a}}, {Freese},
  {Frisch}, {Frolov}, {Fuller}, {Fuzia}, {Galitzki}, {Gallardo}, {Tomas Galvez
  Ghersi}, {Gao}, {Gawiser}, {Gerbino}, {Gluscevic}, {Goeckner-Wald}, {Golec},
  {Gordon}, {Gralla}, {Green}, {Grigorian}, {Groh}, {Groppi}, {Guan},
  {Gudmundsson}, {Han}, {Hargrave}, {Hasegawa}, {Hasselfield}, {Hattori},
  {Haynes}, {Hazumi}, {He}, {Healy}, {Henderson}, {Hervias-Caimapo}, {Hill},
  {Hill}, {Hilton}, {Hilton}, {Hincks}, {Hinshaw}, {Hlo{\v{z}}ek}, {Ho}, {Ho},
  {Howe}, {Huang}, {Hubmayr}, {Huffenberger}, {Hughes}, {Ijjas}, {Ikape},
  {Irwin}, {Jaffe}, {Jain}, {Jeong}, {Kaneko}, {Karpel}, {Katayama}, {Keating},
  {Kernasovskiy}, {Keskitalo}, {Kisner}, {Kiuchi}, {Klein}, {Knowles},
  {Koopman}, {Kosowsky}, {Krachmalnicoff}, {Kuenstner}, {Kuo}, {Kusaka},
  {Lashner}, {Lee}, {Lee}, {Leon}, {Leung}, {Lewis}, {Li}, {Li}, {Limon},
  {Linder}, {Lopez-Caraballo}, {Louis}, {Lowry}, {Lungu}, {Madhavacheril},
  {Mak}, {Maldonado}, {Mani}, {Mates}, {Matsuda}, {Maurin}, {Mauskopf}, {May},
  {McCallum}, {McKenney}, {McMahon}, {Meerburg}, {Meyers}, {Miller},
  {Mirmelstein}, {Moodley}, {Munchmeyer}, {Munson}, {Naess}, {Nati},
  {Navaroli}, {Newburgh}, {Nguyen}, {Niemack}, {Nishino}, {Orlowski-Scherer},
  {Page}, {Partridge}, {Peloton}, {Perrotta}, {Piccirillo}, {Pisano},
  {Poletti}, {Puddu}, {Puglisi}, {Raum}, {Reichardt}, {Remazeilles},
  {Rephaeli}, {Riechers}, {Rojas}, {Roy}, {Sadeh}, {Sakurai}, {Salatino},
  {Sathyanarayana Rao}, {Schaan}, {Schmittfull}, {Sehgal}, {Seibert}, {Seljak},
  {Sherwin}, {Shimon}, {Sierra}, {Sievers}, {Sikhosana}, {Silva-Feaver},
  {Simon}, {Sinclair}, {Siritanasak}, {Smith}, {Smith}, {Spergel}, {Staggs},
  {Stein}, {Stevens}, {Stompor}, {Suzuki}, {Tajima}, {Takakura}, {Teply},
  {Thomas}, {Thorne}, {Thornton}, {Trac}, {Tsai}, {Tucker}, {Ullom},
  {Vagnozzi}, {van Engelen}, {Van Lanen}, {Van Winkle}, {Vavagiakis},
  {Verg{\`e}s}, {Vissers}, {Wagoner}, {Walker}, {Ward}, {Westbrook},
  {Whitehorn}, {Williams}, {Williams}, {Wollack}, {Xu}, {Yu}, {Yu}, {Zago},
  {Zhang}, {Zhu}, \& {Simons Observatory Collaboration}}]{SO2019}
{Ade}, P., {Aguirre}, J., {Ahmed}, Z., {et~al.} 2019, \jcap, 2019, 056

\bibitem[{{Ade} {et~al.}(2021){Ade}, {Ahmed}, {Amiri}, {Barkats}, {Thakur},
  {Bischoff}, {Beck}, {Bock}, {Boenish}, {Bullock}, {Buza}, {Cheshire},
  {Connors}, {Cornelison}, {Crumrine}, {Cukierman}, {Denison}, {Dierickx},
  {Duband}, {Eiben}, {Fatigoni}, {Filippini}, {Fliescher}, {Goeckner-Wald},
  {Goldfinger}, {Grayson}, {Grimes}, {Hall}, {Halal}, {Halpern}, {Hand},
  {Harrison}, {Henderson}, {Hildebrandt}, {Hilton}, {Hubmayr}, {Hui}, {Irwin},
  {Kang}, {Karkare}, {Karpel}, {Kefeli}, {Kernasovskiy}, {Kovac}, {Kuo}, {Lau},
  {Leitch}, {Lennox}, {Megerian}, {Minutolo}, {Moncelsi}, {Nakato}, {Namikawa},
  {Nguyen}, {O'Brient}, {Ogburn}, {Palladino}, {Prouve}, {Pryke}, {Racine},
  {Reintsema}, {Richter}, {Schillaci}, {Schwarz}, {Schmitt}, {Sheehy},
  {Soliman}, {Germaine}, {Steinbach}, {Sudiwala}, {Teply}, {Thompson}, {Tolan},
  {Tucker}, {Turner}, {Umilt{\`a}}, {Verg{\`e}s}, {Vieregg}, {Wandui}, {Weber},
  {Wiebe}, {Willmert}, {Wong}, {Wu}, {Yang}, {Yoon}, {Young}, {Yu}, {Zeng},
  {Zhang}, {Zhang}, \& {Bicep/Keck Collaboration}}]{bicep2021}
{Ade}, P.~A.~R., {Ahmed}, Z., {Amiri}, M., {et~al.} 2021, \prl, 127, 151301

\bibitem[{{Alonso} {et~al.}(2019){Alonso}, {Sanchez}, {Slosar}, \& {LSST Dark
  Energy Science Collaboration}}]{namaster}
{Alonso}, D., {Sanchez}, J., {Slosar}, A., \& {LSST Dark Energy Science
  Collaboration}. 2019, \mnras, 484, 4127

\bibitem[{{Astropy Collaboration} {et~al.}(2022){Astropy Collaboration},
  {Price-Whelan}, {Lim}, {Earl}, {Starkman}, {Bradley}, {Shupe}, {Patil},
  {Corrales}, {Brasseur}, {N{"o}the}, {Donath}, {Tollerud}, {Morris},
  {Ginsburg}, {Vaher}, {Weaver}, {Tocknell}, {Jamieson}, {van Kerkwijk},
  {Robitaille}, {Merry}, {Bachetti}, {G{"u}nther}, {Aldcroft},
  {Alvarado-Montes}, {Archibald}, {B{'o}di}, {Bapat}, {Barentsen}, {Baz{'a}n},
  {Biswas}, {Boquien}, {Burke}, {Cara}, {Cara}, {Conroy}, {Conseil}, {Craig},
  {Cross}, {Cruz}, {D'Eugenio}, {Dencheva}, {Devillepoix}, {Dietrich},
  {Eigenbrot}, {Erben}, {Ferreira}, {Foreman-Mackey}, {Fox}, {Freij}, {Garg},
  {Geda}, {Glattly}, {Gondhalekar}, {Gordon}, {Grant}, {Greenfield}, {Groener},
  {Guest}, {Gurovich}, {Handberg}, {Hart}, {Hatfield-Dodds}, {Homeier},
  {Hosseinzadeh}, {Jenness}, {Jones}, {Joseph}, {Kalmbach}, {Karamehmetoglu},
  {Ka{l}uszy{'n}ski}, {Kelley}, {Kern}, {Kerzendorf}, {Koch}, {Kulumani},
  {Lee}, {Ly}, {Ma}, {MacBride}, {Maljaars}, {Muna}, {Murphy}, {Norman},
  {O'Steen}, {Oman}, {Pacifici}, {Pascual}, {Pascual-Granado}, {Patil},
  {Perren}, {Pickering}, {Rastogi}, {Roulston}, {Ryan}, {Rykoff}, {Sabater},
  {Sakurikar}, {Salgado}, {Sanghi}, {Saunders}, {Savchenko}, {Schwardt},
  {Seifert-Eckert}, {Shih}, {Jain}, {Shukla}, {Sick}, {Simpson},
  {Singanamalla}, {Singer}, {Singhal}, {Sinha}, {Sip{H{o}}cz}, {Spitler},
  {Stansby}, {Streicher}, {{{S}}umak}, {Swinbank}, {Taranu}, {Tewary},
  {Tremblay}, {Val-Borro}, {Van Kooten}, {Vasovi{'c}}, {Verma}, {de Miranda
  Cardoso}, {Williams}, {Wilson}, {Winkel}, {Wood-Vasey}, {Xue}, {Yoachim},
  {Zhang}, {Zonca}, \& {Astropy Project Contributors}}]{astropy:2022}
{Astropy Collaboration}, {Price-Whelan}, A.~M., {Lim}, P.~L., {et~al.} 2022,
  \apj, 935, 167

\bibitem[{{Astropy Collaboration} {et~al.}(2018){Astropy Collaboration},
  {Price-Whelan}, {Sip{\H{o}}cz}, {G{\"u}nther}, {Lim}, {Crawford}, {Conseil},
  {Shupe}, {Craig}, {Dencheva}, {Ginsburg}, {Vand erPlas}, {Bradley},
  {P{\'e}rez-Su{\'a}rez}, {de Val-Borro}, {Aldcroft}, {Cruz}, {Robitaille},
  {Tollerud}, {Ardelean}, {Babej}, {Bach}, {Bachetti}, {Bakanov}, {Bamford},
  {Barentsen}, {Barmby}, {Baumbach}, {Berry}, {Biscani}, {Boquien}, {Bostroem},
  {Bouma}, {Brammer}, {Bray}, {Breytenbach}, {Buddelmeijer}, {Burke},
  {Calderone}, {Cano Rodr{\'\i}guez}, {Cara}, {Cardoso}, {Cheedella}, {Copin},
  {Corrales}, {Crichton}, {D'Avella}, {Deil}, {Depagne}, {Dietrich}, {Donath},
  {Droettboom}, {Earl}, {Erben}, {Fabbro}, {Ferreira}, {Finethy}, {Fox},
  {Garrison}, {Gibbons}, {Goldstein}, {Gommers}, {Greco}, {Greenfield},
  {Groener}, {Grollier}, {Hagen}, {Hirst}, {Homeier}, {Horton}, {Hosseinzadeh},
  {Hu}, {Hunkeler}, {Ivezi{\'c}}, {Jain}, {Jenness}, {Kanarek}, {Kendrew},
  {Kern}, {Kerzendorf}, {Khvalko}, {King}, {Kirkby}, {Kulkarni}, {Kumar},
  {Lee}, {Lenz}, {Littlefair}, {Ma}, {Macleod}, {Mastropietro}, {McCully},
  {Montagnac}, {Morris}, {Mueller}, {Mumford}, {Muna}, {Murphy}, {Nelson},
  {Nguyen}, {Ninan}, {N{\"o}the}, {Ogaz}, {Oh}, {Parejko}, {Parley}, {Pascual},
  {Patil}, {Patil}, {Plunkett}, {Prochaska}, {Rastogi}, {Reddy Janga},
  {Sabater}, {Sakurikar}, {Seifert}, {Sherbert}, {Sherwood-Taylor}, {Shih},
  {Sick}, {Silbiger}, {Singanamalla}, {Singer}, {Sladen}, {Sooley},
  {Sornarajah}, {Streicher}, {Teuben}, {Thomas}, {Tremblay}, {Turner},
  {Terr{\'o}n}, {van Kerkwijk}, {de la Vega}, {Watkins}, {Weaver}, {Whitmore},
  {Woillez}, {Zabalza}, \& {Astropy Contributors}}]{astropy:2018}
{Astropy Collaboration}, {Price-Whelan}, A.~M., {Sip{\H{o}}cz}, B.~M., {et~al.}
  2018, \aj, 156, 123

\bibitem[{{Astropy Collaboration} {et~al.}(2013){Astropy Collaboration},
  {Robitaille}, {Tollerud}, {Greenfield}, {Droettboom}, {Bray}, {Aldcroft},
  {Davis}, {Ginsburg}, {Price-Whelan}, {Kerzendorf}, {Conley}, {Crighton},
  {Barbary}, {Muna}, {Ferguson}, {Grollier}, {Parikh}, {Nair}, {Unther},
  {Deil}, {Woillez}, {Conseil}, {Kramer}, {Turner}, {Singer}, {Fox}, {Weaver},
  {Zabalza}, {Edwards}, {Azalee Bostroem}, {Burke}, {Casey}, {Crawford},
  {Dencheva}, {Ely}, {Jenness}, {Labrie}, {Lim}, {Pierfederici}, {Pontzen},
  {Ptak}, {Refsdal}, {Servillat}, \& {Streicher}}]{astropy:2013}
{Astropy Collaboration}, {Robitaille}, T.~P., {Tollerud}, E.~J., {et~al.} 2013,
  \aap, 558, A33

\bibitem[{{Bennett} {et~al.}(1996){Bennett}, {Banday}, {Gorski}, {Hinshaw},
  {Jackson}, {Keegstra}, {Kogut}, {Smoot}, {Wilkinson}, \& {Wright}}]{dmr}
{Bennett}, C.~L., {Banday}, A.~J., {Gorski}, K.~M., {et~al.} 1996, \apjl, 464,
  L1

\bibitem[{{Bennett} {et~al.}(2013){Bennett}, {Larson}, {Weiland}, {Jarosik},
  {Hinshaw}, {Odegard}, {Smith}, {Hill}, {Gold}, {Halpern}, {Komatsu}, {Nolta},
  {Page}, {Spergel}, {Wollack}, {Dunkley}, {Kogut}, {Limon}, {Meyer}, {Tucker},
  \& {Wright}}]{bennett2012}
{Bennett}, C.~L., {Larson}, D., {Weiland}, J.~L., {et~al.} 2013, \apjs, 208, 20

\bibitem[{{BeyondPlanck Collaboration}(2023)}]{bp01}
{BeyondPlanck Collaboration}. 2023, A\&A, 675, A1

\bibitem[{{BICEP2/Keck Array and Planck Collaborations}(2015)}]{pb2015}
{BICEP2/Keck Array and Planck Collaborations}. 2015, \prl, 114, 101301

\bibitem[{{Carlstrom} {et~al.}(2011){Carlstrom}, {Ade}, {Aird}, {Benson},
  {Bleem}, {Busetti}, {Chang}, {Chauvin}, {Cho}, {Crawford}, {Crites}, {Dobbs},
  {Halverson}, {Heimsath}, {Holzapfel}, {Hrubes}, {Joy}, {Keisler}, {Lanting},
  {Lee}, {Leitch}, {Leong}, {Lu}, {Lueker}, {Luong-Van}, {McMahon}, {Mehl},
  {Meyer}, {Mohr}, {Montroy}, {Padin}, {Plagge}, {Pryke}, {Ruhl}, {Schaffer},
  {Schwan}, {Shirokoff}, {Spieler}, {Staniszewski}, {Stark}, {Tucker},
  {Vanderlinde}, {Vieira}, \& {Williamson}}]{carlstrom:2011}
{Carlstrom}, J.~E., {Ade}, P.~A.~R., {Aird}, K.~A., {et~al.} 2011, \pasp, 123,
  568

\bibitem[{{de Belsunce} {et~al.}(2022){de Belsunce}, {Gratton}, \&
  {Efstathiou}}]{deBelsunce:2022}
{de Belsunce}, R., {Gratton}, S., \& {Efstathiou}, G. 2022, \mnras, 517, 2855

\bibitem[{{de la Hoz} {et~al.}(2023){de la Hoz}, {Barreiro}, {Vielva},
  {Mart{\'\i}nez-Gonz{\'a}lez}, {Rubi{\~n}o-Mart{\'\i}n}, {Casaponsa}, {Guidi},
  {Ashdown}, {G{\'e}nova-Santos}, {Artal}, {Casas}, {Fern{\'a}ndez-Cobos},
  {Fern{\'a}ndez-Torreiro}, {Herranz}, {Hoyland}, {Lasenby},
  {L{\'o}pez-Caniego}, {L{\'o}pez-Caraballo}, {Peel}, {Piccirillo}, {Poidevin},
  {Rebolo}, {Ruiz-Granados}, {Tramonte}, {Vansyngel}, \&
  {Watson}}]{QUIJOTE_VIII}
{de la Hoz}, E., {Barreiro}, R.~B., {Vielva}, P., {et~al.} 2023, \mnras, 519,
  3504

\bibitem[{{de la Hoz} {et~al.}(2022){de la Hoz}, {Diego-Palazuelos},
  {Mart{\'\i}nez-Gonz{\'a}lez}, {Vielva}, {Barreiro}, \&
  {Bilbao-Ahedo}}]{b-secret}
{de la Hoz}, E., {Diego-Palazuelos}, P., {Mart{\'\i}nez-Gonz{\'a}lez}, E.,
  {et~al.} 2022, \jcap, 2022, 032

\bibitem[{{Delabrouille} {et~al.}(2013){Delabrouille}, {Betoule}, {Melin},
  {Miville-Desch{\^e}nes}, {Gonzalez-Nuevo}, {Le Jeune}, {Castex}, {de Zotti},
  {Basak}, {Ashdown}, {Aumont}, {Baccigalupi}, {Banday}, {Bernard}, {Bouchet},
  {Clements}, {da Silva}, {Dickinson}, {Dodu}, {Dolag}, {Elsner}, {Fauvet},
  {Fa{\"y}}, {Giardino}, {Leach}, {Lesgourgues}, {Liguori}, {Macias-Perez},
  {Massardi}, {Matarrese}, {Mazzotta}, {Montier}, {Mottet}, {Paladini},
  {Partridge}, {Piffaretti}, {Prezeau}, {Prunet}, {Ricciardi}, {Roman},
  {Schaefer}, \& {Toffolatti}}]{delabrouille2012}
{Delabrouille}, J., {Betoule}, M., {Melin}, J.-B., {et~al.} 2013, \aap, 553,
  A96

\bibitem[{{Eimer} {et~al.}(2023){Eimer}, {Li}, {Brewer}, {Shi}, {Ali}, {Appel},
  {Bennett}, {Bustos}, {Chuss}, {Cleary}, {Dahal}, {Datta}, {Denes Couto},
  {Denis}, {D{\"u}nner}, {Essinger-Hileman}, {Flux{\'a}}, {Hubmayer},
  {Harrington}, {Iuliano}, {Karakla}, {Marriage}, {N{\'u}{\~n}ez}, {Parker},
  {Petroff}, {Reeves}, {Rostem}, {Valle}, {Watts}, {Weiland}, {Wollack}, {Xu},
  \& {Zeng}}]{eimer2023}
{Eimer}, J.~R., {Li}, Y., {Brewer}, M.~K., {et~al.} 2023, arXiv e-prints,
  arXiv:2309.00675

\bibitem[{{Eriksen} {et~al.}(2008){Eriksen}, {Jewell}, {Dickinson}, {Banday},
  {G{\'o}rski}, \& {Lawrence}}]{eriksen2008}
{Eriksen}, H.~K., {Jewell}, J.~B., {Dickinson}, C., {et~al.} 2008, \apj, 676,
  10

\bibitem[{{Fuskeland} {et~al.}(2021){Fuskeland}, {Andersen}, {Aurlien},
  {Banerji}, {Brilenkov}, {Eriksen}, {Galloway}, {Gjerl{\o}w}, {N{\ae}ss},
  {Svalheim}, \& {Wehus}}]{fuskeland:2019}
{Fuskeland}, U., {Andersen}, K.~J., {Aurlien}, R., {et~al.} 2021, \aap, 646,
  A69

\bibitem[{{Fuskeland} {et~al.}(2014){Fuskeland}, {Wehus}, {Eriksen}, \&
  {N{\ae}ss}}]{fuskeland2014}
{Fuskeland}, U., {Wehus}, I.~K., {Eriksen}, H.~K., \& {N{\ae}ss}, S.~K. 2014,
  \apj, 790, 104

\bibitem[{{Galloway} {et~al.}(2023){Galloway}, {Andersen, K. J.}, {Aurlien,
  R.}, {Banerji, R.}, {Bersanelli, M.}, {Bertocco, S.}, {Brilenkov, M.},
  {Carbone, M.}, {Colombo, L. P. L.}, {Eriksen, H. K.}, {Eskilt, J. R.}, {Foss,
  M. K.}, {Franceschet, C.}, {Fuskeland, U.}, {Galeotta, S.}, {Gerakakis, S.},
  {Gjerl\o{}w, E.}, {Hensley, B.}, {Herman, D.}, {Iacobellis, M.}, {Ieronymaki,
  M.}, {Ihle, H. T.}, {Jewell, J. B.}, {Karakci, A.}, {Keih\"anen, E.},
  {Keskitalo, R.}, {Maggio, G.}, {Maino, D.}, {Maris, M.}, {Mennella, A.},
  {Paradiso, S.}, {Partridge, B.}, {Reinecke, M.}, {San, M.}, {Suur-Uski,
  A.-S.}, {Svalheim, T. L.}, {Tavagnacco, D.}, {Thommesen, H.}, {Watts, D. J.},
  {Wehus, I. K.}, \& {Zacchei, A.}}]{bp03}
{Galloway}, M., {Andersen, K. J.}, {Aurlien, R.}, {et~al.} 2023, A\&A, 675, A3

\bibitem[{{Gjerl\o{}w} {et~al.}(2023){Gjerl\o{}w}, {Ihle, H. T.}, {Galeotta,
  S.}, {Andersen, K. J.}, {Aurlien, R.}, {Banerji, R.}, {Bersanelli, M.},
  {Bertocco, S.}, {Brilenkov, M.}, {Carbone, M.}, {Colombo, L. P. L.},
  {Eriksen, H. K.}, {Foss, M. K.}, {Franceschet, C.}, {Fuskeland, U.},
  {Galloway, M.}, {Gerakakis, S.}, {Hensley, B.}, {Herman, D.}, {Iacobellis,
  M.}, {Ieronymaki, M.}, {Jewell, J. B.}, {Karakci, A.}, {Keih\"anen, E.},
  {Keskitalo, R.}, {Maggio, G.}, {Maino, D.}, {Maris, M.}, {Paradiso, S.},
  {Partridge, B.}, {Reinecke, M.}, {Suur-Uski, A.-S.}, {Svalheim, T. L.},
  {Tavagnacco, D.}, {Thommesen, H.}, {Watts, D. J.}, {Wehus, I. K.}, \&
  {Zacchei, A.}}]{bp07}
{Gjerl\o{}w}, E., {Ihle, H. T.}, {Galeotta, S.}, {et~al.} 2023, A\&A, 675, A7

\bibitem[{{G{\'o}rski} {et~al.}(2005){G{\'o}rski}, {Hivon}, {Banday},
  {Wandelt}, {Hansen}, {Reinecke}, \& {Bartelmann}}]{gorski2005}
{G{\'o}rski}, K.~M., {Hivon}, E., {Banday}, A.~J., {et~al.} 2005, \apj, 622,
  759

\bibitem[{{Haslam} {et~al.}(1982){Haslam}, {Salter}, {Stoffel}, \&
  {Wilson}}]{haslam1982}
{Haslam}, C.~G.~T., {Salter}, C.~J., {Stoffel}, H., \& {Wilson}, W.~E. 1982,
  \aaps, 47, 1

\bibitem[{{Hauser} {et~al.}(1998){Hauser}, {Arendt}, {Kelsall}, {Dwek},
  {Odegard}, {Weiland}, {Freudenreich}, {Reach}, {Silverberg}, {Moseley},
  {Pei}, {Lubin}, {Mather}, {Shafer}, {Smoot}, {Weiss}, {Wilkinson}, \&
  {Wright}}]{hauser:1998}
{Hauser}, M.~G., {Arendt}, R.~G., {Kelsall}, T., {et~al.} 1998, \apj, 508, 25

\bibitem[{{Hensley} {et~al.}(2022){Hensley}, {Clark}, {Fanfani},
  {Krachmalnicoff}, {Fabbian}, {Poletti}, {Puglisi}, {Coppi}, {Nibauer},
  {Gerasimov}, {Galitzki}, {Choi}, {Ashton}, {Baccigalupi}, {Baxter},
  {Burkhart}, {Calabrese}, {Chluba}, {Errard}, {Frolov},
  {Herv{\'\i}as-Caimapo}, {Huffenberger}, {Johnson}, {Jost}, {Keating},
  {McCarrick}, {Nati}, {Sathyanarayana Rao}, {van Engelen}, {Walker}, {Wolz},
  {Xu}, {Zhu}, \& {Zonca}}]{so_galsci}
{Hensley}, B.~S., {Clark}, S.~E., {Fanfani}, V., {et~al.} 2022, \apj, 929, 166

\bibitem[{{Jarosik} {et~al.}(2007){Jarosik}, {Barnes}, {Greason}, {Hill},
  {Nolta}, {Odegard}, {Weiland}, {Bean}, {Bennett}, {Dor{\'e}}, {Halpern},
  {Hinshaw}, {Kogut}, {Komatsu}, {Limon}, {Meyer}, {Page}, {Spergel}, {Tucker},
  {Wollack}, \& {Wright}}]{jarosik2007}
{Jarosik}, N., {Barnes}, C., {Greason}, M.~R., {et~al.} 2007, \apjs, 170, 263

\bibitem[{{Jarosik} {et~al.}(2003){Jarosik}, {Bennett}, {Halpern}, {Hinshaw},
  {Kogut}, {Limon}, {Meyer}, {Page}, {Pospieszalski}, {Spergel}, {Tucker},
  {Wilkinson}, {Wollack}, {Wright}, \& {Zhang}}]{jarosik2003:MAP}
{Jarosik}, N., {Bennett}, C.~L., {Halpern}, M., {et~al.} 2003, \apjs, 145, 413

\bibitem[{{Kandel} {et~al.}(2017){Kandel}, {Lazarian}, \&
  {Pogosyan}}]{kandel2017}
{Kandel}, D., {Lazarian}, A., \& {Pogosyan}, D. 2017, \mnras, 472, L10

\bibitem[{{Krachmalnicoff} {et~al.}(2018){Krachmalnicoff}, {Carretti},
  {Baccigalupi}, {Bernardi}, {Brown}, {Gaensler}, {Haverkorn}, {Kesteven},
  {Perrotta}, {Poppi}, \& {Staveley-Smith}}]{krachmalnicoff2018}
{Krachmalnicoff}, N., {Carretti}, E., {Baccigalupi}, C., {et~al.} 2018, \aap,
  618, A166

\bibitem[{{LiteBIRD Collaboration} {et~al.}(2023){LiteBIRD Collaboration},
  {Allys}, {Arnold}, {Aumont}, {Aurlien}, {Azzoni}, {Baccigalupi}, {Banday},
  {Banerji}, {Barreiro}, {Bartolo}, {Bautista}, {Beck}, {Beckman},
  {Bersanelli}, {Boulanger}, {Brilenkov}, {Bucher}, {Calabrese}, {Campeti},
  {Carones}, {Casas}, {Catalano}, {Chan}, {Cheung}, {Chinone}, {Clark},
  {Columbro}, {D'Alessandro}, {de Bernardis}, {de Haan}, {de la Hoz}, {De
  Petris}, {Torre}, {Diego-Palazuelos}, {Dobbs}, {Dotani}, {Duval}, {Elleflot},
  {Eriksen}, {Errard}, {Essinger-Hileman}, {Finelli}, {Flauger}, {Franceschet},
  {Fuskeland}, {Galloway}, {Ganga}, {Gerbino}, {Gervasi}, {G{\'e}nova-Santos},
  {Ghigna}, {Giardiello}, {Gjerl{\o}w}, {Grain}, {Grupp}, {Gruppuso},
  {Gudmundsson}, {Halverson}, {Hargrave}, {Hasebe}, {Hasegawa}, {Hazumi},
  {Henrot-Versill{\'e}}, {Hensley}, {Hergt}, {Herman}, {Hivon}, {Hlozek},
  {Hornsby}, {Hoshino}, {Hubmayr}, {Ichiki}, {Iida}, {Imada}, {Ishino},
  {Jaehnig}, {Katayama}, {Kato}, {Keskitalo}, {Kisner}, {Kobayashi}, {Kogut},
  {Kohri}, {Komatsu}, {Komatsu}, {Konishi}, {Krachmalnicoff}, {Kuo}, {Lamagna},
  {Lattanzi}, {Lee}, {Leloup}, {Levrier}, {Linder}, {Luzzi}, {Macias-Perez},
  {Maciaszek}, {Maffei}, {Maino}, {Mandelli}, {Mart{\'\i}nez-Gonz{\'a}lez},
  {Masi}, {Massa}, {Matarrese}, {Matsuda}, {Matsumura}, {Mele}, {Migliaccio},
  {Minami}, {Moggi}, {Montgomery}, {Montier}, {Morgante}, {Mot}, {Nagano},
  {Nagasaki}, {Nagata}, {Nakano}, {Namikawa}, {Nati}, {Natoli}, {Nerval},
  {Noviello}, {Odagiri}, {Oguri}, {Ohsaki}, {Pagano}, {Paiella}, {Paoletti},
  {Passerini}, {Patanchon}, {Piacentini}, {Piat}, {Pisano}, {Polenta},
  {Poletti}, {Prouv{\'e}}, {Puglisi}, {Rambaud}, {Raum}, {Realini}, {Reinecke},
  {Remazeilles}, {Ritacco}, {Roudil}, {Rubino-Martin}, {Russell}, {Sakurai},
  {Sakurai}, {Sasaki}, {Scott}, {Sekimoto}, {Shinozaki}, {Shiraishi},
  {Shirron}, {Signorelli}, {Spinella}, {Stever}, {Stompor}, {Sugiyama},
  {Sullivan}, {Suzuki}, {Svalheim}, {Switzer}, {Takaku}, {Takakura}, {Takase},
  {Tartari}, {Terao}, {Thermeau}, {Thommesen}, {Thompson}, {Tomasi},
  {Tominaga}, {Tristram}, {Tsuji}, {Tsujimoto}, {Vacher}, {Vielva}, {Vittorio},
  {Wang}, {Watanuki}, {Wehus}, {Weller}, {Westbrook}, {Wilms}, {Winter},
  {Wollack}, {Yumoto}, {Zannoni}, \& {Collaboration LiteB I R
  D}}]{litebird2022}
{LiteBIRD Collaboration}, {Allys}, E., {Arnold}, K., {et~al.} 2023, Progress of
  Theoretical and Experimental Physics, 2023, 042F01

\bibitem[{{Madhavacheril} {et~al.}(2023){Madhavacheril}, {Qu}, {Sherwin},
  {MacCrann}, {Li}, {Abril-Cabezas}, {Ade}, {Aiola}, {Alford}, {Amiri},
  {Amodeo}, {An}, {Atkins}, {Austermann}, {Battaglia}, {Battistelli}, {Beall},
  {Bean}, {Beringue}, {Bhandarkar}, {Biermann}, {Bolliet}, {Bond}, {Cai},
  {Calabrese}, {Calafut}, {Capalbo}, {Carrero}, {Challinor}, {Chesmore}, {Cho},
  {Choi}, {Clark}, {C{\'o}rdova Rosado}, {Cothard}, {Coughlin}, {Coulton},
  {Crowley}, {Dalal}, {Darwish}, {Devlin}, {Dicker}, {Doze}, {Duell}, {Duff},
  {Duivenvoorden}, {Dunkley}, {D{\"u}nner}, {Fanfani}, {Fankhanel}, {Farren},
  {Ferraro}, {Freundt}, {Fuzia}, {Gallardo}, {Garrido}, {Givans}, {Gluscevic},
  {Golec}, {Guan}, {Hall}, {Halpern}, {Han}, {Harrison}, {Hasselfield},
  {Healy}, {Henderson}, {Hensley}, {Herv{\'\i}as-Caimapo}, {Hill}, {Hilton},
  {Hilton}, {Hincks}, {Hlo{\v{z}}ek}, {Ho}, {Huber}, {Hubmayr}, {Huffenberger},
  {Hughes}, {Irwin}, {Isopi}, {Jense}, {Keller}, {Kim}, {Knowles}, {Koopman},
  {Kosowsky}, {Kramer}, {Kusiak}, {La Posta}, {Lague}, {Lakey}, {Lee}, {Li},
  {Limon}, {Lokken}, {Louis}, {Lungu}, {MacInnis}, {Maldonado}, {Maldonado},
  {Mallaby-Kay}, {Marques}, {McMahon}, {Mehta}, {Menanteau}, {Moodley},
  {Morris}, {Mroczkowski}, {Naess}, {Namikawa}, {Nati}, {Newburgh}, {Nicola},
  {Niemack}, {Nolta}, {Orlowski-Scherer}, {Page}, {Pandey}, {Partridge},
  {Prince}, {Puddu}, {Radiconi}, {Robertson}, {Rojas}, {Sakuma}, {Salatino},
  {Schaan}, {Schmitt}, {Sehgal}, {Shaikh}, {Sierra}, {Sievers}, {Sif{\'o}n},
  {Simon}, {Sonka}, {Spergel}, {Staggs}, {Storer}, {Switzer}, {Tampier},
  {Thornton}, {Trac}, {Treu}, {Tucker}, {Ulluom}, {Vale}, {Van Engelen}, {Van
  Lanen}, {van Marrewijk}, {Vargas}, {Vavagiakis}, {Wagoner}, {Wang}, {Wenzl},
  {Wollack}, {Xu}, {Zago}, \& {Zhang}}]{actDR6_lensing}
{Madhavacheril}, M.~S., {Qu}, F.~J., {Sherwin}, B.~D., {et~al.} 2023, arXiv
  e-prints, arXiv:2304.05203

\bibitem[{{Mather} {et~al.}(1994){Mather}, {Cheng}, {Cottingham}, {Eplee},
  {Fixsen}, {Hewagama}, {Isaacman}, {Jensen}, {Meyer}, {Noerdlinger}, {Read},
  {Rosen}, {Shafer}, {Wright}, {Bennett}, {Boggess}, {Hauser}, {Kelsall},
  {Moseley}, {Silverberg}, {Smoot}, {Weiss}, \& {Wilkinson}}]{mather:1994}
{Mather}, J.~C., {Cheng}, E.~S., {Cottingham}, D.~A., {et~al.} 1994, \apj, 420,
  439

\bibitem[{{Neronov} {et~al.}(2017){Neronov}, {Malyshev}, \&
  {Semikoz}}]{neronov2017}
{Neronov}, A., {Malyshev}, D., \& {Semikoz}, D.~V. 2017, \aap, 606, A22

\bibitem[{{Orear}(1982)}]{orear1982}
{Orear}, J. 1982, American Journal of Physics, 50, 912

\bibitem[{{Orlando} \& {Strong}(2013)}]{orlando2013}
{Orlando}, E. \& {Strong}, A. 2013, \mnras, 436, 2127

\bibitem[{{Page} {et~al.}(2007){Page}, {Hinshaw}, {Komatsu}, {Nolta},
  {Spergel}, {Bennett}, {Barnes}, {Bean}, {Dor{\'e}}, {Dunkley}, {Halpern},
  {Hill}, {Jarosik}, {Kogut}, {Limon}, {Meyer}, {Odegard}, {Peiris}, {Tucker},
  {Verde}, {Weiland}, {Wollack}, \& {Wright}}]{page2007}
{Page}, L., {Hinshaw}, G., {Komatsu}, E., {et~al.} 2007, \apjs, 170, 335

\bibitem[{{\sorthelp{Planck Collaboration 2015J}}{Planck Collaboration
  X}(2016)}]{planck2014-a12}
{\sorthelp{Planck Collaboration 2015J}}{Planck Collaboration X}. 2016, \aap,
  594, A10

\bibitem[{{\sorthelp{Planck Collaboration 2018A}}{Planck Collaboration
  I}(2020)}]{planck2016-l01}
{\sorthelp{Planck Collaboration 2018A}}{Planck Collaboration I}. 2020, \aap,
  641, A1

\bibitem[{{\sorthelp{Planck Collaboration 2018B}}{Planck Collaboration
  II}(2020)}]{planck2016-l02}
{\sorthelp{Planck Collaboration 2018B}}{Planck Collaboration II}. 2020, \aap,
  641, A2

\bibitem[{{\sorthelp{Planck Collaboration 2018D}}{Planck Collaboration
  IV}(2018)}]{planck2016-l04}
{\sorthelp{Planck Collaboration 2018D}}{Planck Collaboration IV}. 2018, \aap,
  641, A4

\bibitem[{{\sorthelp{Planck Collaboration 2018E}}{Planck Collaboration
  V}(2020)}]{planck2016-l05}
{\sorthelp{Planck Collaboration 2018E}}{Planck Collaboration V}. 2020, \aap,
  641, A5

\bibitem[{{\sorthelp{Planck Collaboration IntZZG}}{Planck Collaboration Int.
  LVII}(2020)}]{planck2020-LVII}
{\sorthelp{Planck Collaboration IntZZG}}{Planck Collaboration Int. LVII}. 2020,
  \aap, 643, A42

\bibitem[{{Rubi{\~n}o-Mart{\'\i}n} {et~al.}(2023){Rubi{\~n}o-Mart{\'\i}n},
  {Guidi}, {G{\'e}nova-Santos}, {Harper}, {Herranz}, {Hoyland}, {Lasenby},
  {Poidevin}, {Rebolo}, {Ruiz-Granados}, {Vansyngel}, {Vielva}, {Watson},
  {Artal}, {Ashdown}, {Barreiro}, {Bilbao-Ahedo}, {Casas}, {Casaponsa},
  {Cepeda-Arroita}, {de la Hoz}, {Dickinson}, {Fern{\'a}ndez-Cobos},
  {Fern{\'a}ndez-Torreiro}, {Gonz{\'a}lez-Gonz{\'a}lez},
  {Hern{\'a}ndez-Monteagudo}, {L{\'o}pez-Caniego}, {L{\'o}pez-Caraballo},
  {Mart{\'\i}nez-Gonz{\'a}lez}, {Peel}, {Pel{\'a}ez-Santos}, {Perrott},
  {Piccirillo}, {Razavi-Ghods}, {Scott}, {Titterington}, {Tramonte}, \&
  {Vignaga}}]{QUIJOTE_IV}
{Rubi{\~n}o-Mart{\'\i}n}, J.~A., {Guidi}, F., {G{\'e}nova-Santos}, R.~T.,
  {et~al.} 2023, \mnras, 519, 3383

\bibitem[{Rybicki \& Lightman(1985)}]{rybicki}
Rybicki, G.~B. \& Lightman, A.~P. 1985, {Radiative Processes in Astrophysics}
  (New York, NY: Wiley)

\bibitem[{{Smoot} {et~al.}(1992){Smoot}, {Bennett}, {Kogut}, {Wright}, {Aymon},
  {Boggess}, {Cheng}, {de Amici}, {Gulkis}, {Hauser}, {Hinshaw}, {Jackson},
  {Janssen}, {Kaita}, {Kelsall}, {Keegstra}, {Lineweaver}, {Loewenstein},
  {Lubin}, {Mather}, {Meyer}, {Moseley}, {Murdock}, {Rokke}, {Silverberg},
  {Tenorio}, {Weiss}, \& {Wilkinson}}]{smoot:1992}
{Smoot}, G.~F., {Bennett}, C.~L., {Kogut}, A., {et~al.} 1992, \apjl, 396, L1

\bibitem[{{SPIDER Collaboration} {et~al.}(2022){SPIDER Collaboration}, {Ade},
  {Amiri}, {Benton}, {Bergman}, {Bihary}, {Bock}, {Bond}, {Bonetti}, {Bryan},
  {Chiang}, {Contaldi}, {Dor{\'e}}, {Duivenvoorden}, {Eriksen}, {Farhang},
  {Filippini}, {Fraisse}, {Freese}, {Galloway}, {Gambrel}, {Gandilo}, {Ganga},
  {Gualtieri}, {Gudmundsson}, {Halpern}, {Hartley}, {Hasselfield}, {Hilton},
  {Holmes}, {Hristov}, {Huang}, {Irwin}, {Jones}, {Karakci}, {Kuo}, {Kermish},
  {Leung}, {Li}, {Mak}, {Mason}, {Megerian}, {Moncelsi}, {Morford}, {Nagy},
  {Netterfield}, {Nolta}, {O'Brient}, {Osherson}, {Padilla}, {Racine},
  {Rahlin}, {Reintsema}, {Ruhl}, {Runyan}, {Ruud}, {Shariff}, {Shaw}, {Shiu},
  {Soler}, {Song}, {Trangsrud}, {Tucker}, {Tucker}, {Turner}, {van der List},
  {Weber}, {Wehus}, {Wen}, {Wiebe}, \& {Young}}]{spider21}
{SPIDER Collaboration}, {Ade}, P.~A.~R., {Amiri}, M., {et~al.} 2022, \apj, 927,
  174

\bibitem[{{Svalheim} {et~al.}(2023{\natexlab{a}}){Svalheim}, {Andersen, K. J.},
  {Aurlien, R.}, {Banerji, R.}, {Bersanelli, M.}, {Bertocco, S.}, {Brilenkov,
  M.}, {Carbone, M.}, {Colombo, L. P. L.}, {Eriksen, H. K.}, {Foss, M. K.},
  {Franceschet, C.}, {Fuskeland, U.}, {Galeotta, S.}, {Galloway, M.},
  {Gerakakis, S.}, {Gjerl\o{}w, E.}, {Hensley, B.}, {Herman, D.}, {Iacobellis,
  M.}, {Ieronymaki, M.}, {Ihle, H. T.}, {Jewell, J. B.}, {Karakci, A.},
  {Keih\"anen, E.}, {Keskitalo, R.}, {Maggio, G.}, {Maino, D.}, {Maris, M.},
  {Paradiso, S.}, {Partridge, B.}, {Reinecke, M.}, {Suur-Uski, A.-S.},
  {Tavagnacco, D.}, {Thommesen, H.}, {Watts, D. J.}, {Wehus, I. K.}, \&
  {Zacchei, A.}}]{bp14}
{Svalheim}, T.~L., {Andersen, K. J.}, {Aurlien, R.}, {et~al.}
  2023{\natexlab{a}}, A\&A, 675, A14

\bibitem[{{Svalheim} {et~al.}(2023{\natexlab{b}}){Svalheim}, {Zonca, A.},
  {Andersen, K. J.}, {Aurlien, R.}, {Banerji, R.}, {Bersanelli, M.}, {Bertocco,
  S.}, {Brilenkov, M.}, {Carbone, M.}, {Colombo, L. P. L.}, {Eriksen, H. K.},
  {Foss, M. K.}, {Franceschet, C.}, {Fuskeland, U.}, {Galeotta, S.}, {Galloway,
  M.}, {Gerakakis, S.}, {Gjerl\o{}w, E.}, {Hensley, B.}, {Herman, D.},
  {Iacobellis, M.}, {Ieronymaki, M.}, {Ihle, H. T.}, {Jewell, J. B.}, {Karakci,
  A.}, {Keih\"anen, E.}, {Keskitalo, R.}, {Maggio, G.}, {Maino, D.}, {Maris,
  M.}, {Paradiso, S.}, {Partridge, B.}, {Reinecke, M.}, {Suur-Uski, A.-S.},
  {Tavagnacco, D.}, {Thommesen, H.}, {Watts, D. J.}, {Wehus, I. K.}, \&
  {Zacchei, A.}}]{bp09}
{Svalheim}, T.~L., {Zonca, A.}, {Andersen, K. J.}, {et~al.} 2023{\natexlab{b}},
  A\&A, 675, A9

\bibitem[{{Thorne} {et~al.}(2017){Thorne}, {Dunkley}, {Alonso}, \&
  {N{\ae}ss}}]{pysm2}
{Thorne}, B., {Dunkley}, J., {Alonso}, D., \& {N{\ae}ss}, S. 2017, \mnras, 469,
  2821

\bibitem[{{Vacher} {et~al.}(2023){Vacher}, {Aumont}, {Boulanger}, {Montier},
  {Guillet}, {Ritacco}, \& {Chluba}}]{vacher2023}
{Vacher}, L., {Aumont}, J., {Boulanger}, F., {et~al.} 2023, \aap, 672, A146

\bibitem[{{Watts} {et~al.}(2023{\natexlab{a}}){Watts}, {Galloway, M.}, {Ihle,
  H. T.}, {Andersen, K. J.}, {Aurlien, R.}, {Banerji, R.}, {Basyrov, A.},
  {Bersanelli, M.}, {Bertocco, S.}, {Brilenkov, M.}, {Carbone, M.}, {Colombo,
  L. P. L.}, {Eriksen, H. K.}, {Eskilt, J. R.}, {Foss, M. K.}, {Franceschet,
  C.}, {Fuskeland, U.}, {Galeotta, S.}, {Gerakakis, S.}, {Gjerl\o{}w, E.},
  {Hensley, B.}, {Herman, D.}, {Iacobellis, M.}, {Ieronymaki, M.}, {Jewell, J.
  B.}, {Karakci, A.}, {Keih\"anen, E.}, {Keskitalo, R.}, {Lunde, J. G. S.},
  {Maggio, G.}, {Maino, D.}, {Maris, M.}, {Paradiso, S.}, {Partridge, B.},
  {Reinecke, M.}, {San, M.}, {Stutzer, N.-O.}, {Suur-Uski, A.-S.}, {Svalheim,
  T. L.}, {Tavagnacco, D.}, {Thommesen, H.}, {Wehus, I. K.}, \& {Zacchei,
  A.}}]{bp17}
{Watts}, D.~J., {Galloway, M.}, {Ihle, H. T.}, {et~al.} 2023{\natexlab{a}},
  A\&A, 675, A16

\bibitem[{{Watts} {et~al.}(2023{\natexlab{b}}){Watts}, {Basyrov}, {Eskilt},
  {Galloway}, {Hergt}, {Herman}, {Ihle}, {Paradiso}, {Rahman}, {Thommesen},
  {Aurlien}, {Bersanelli}, {Bianchi}, {Brilenkov}, {Colombo}, {Eriksen},
  {Franceschet}, {Fuskeland}, {Gjerl{\o}w}, {Hensley}, {Hoerning}, {Lee},
  {Lunde}, {Marins}, {Nerval}, {Patel}, {Regnier}, {San}, {Sanyal}, {Stutzer},
  {Verma}, {Wehus}, \& {Zhou}}]{watts2023_dr1}
{Watts}, D.~J., {Basyrov}, A., {Eskilt}, J.~R., {et~al.} 2023{\natexlab{b}},
  arXiv e-prints, arXiv:2303.08095

\bibitem[{{Wehus} {et~al.}(2013){Wehus}, {Fuskeland}, \&
  {Eriksen}}]{wehus:2013}
{Wehus}, I.~K., {Fuskeland}, U., \& {Eriksen}, H.~K. 2013, \apj, 763, 138

\bibitem[{{Weiland} {et~al.}(2022){Weiland}, {Addison}, {Bennett}, {Halpern},
  \& {Hinshaw}}]{weiland:2022}
{Weiland}, J.~L., {Addison}, G.~E., {Bennett}, C.~L., {Halpern}, M., \&
  {Hinshaw}, G. 2022, \apj, 936, 24

\bibitem[{{Weiland} {et~al.}(2018){Weiland}, {Osumi}, {Addison}, {Bennett},
  {Watts}, {Halpern}, \& {Hinshaw}}]{weiland:2018}
{Weiland}, J.~L., {Osumi}, K., {Addison}, G.~E., {et~al.} 2018, \apj, 863, 161

\bibitem[{Zonca {et~al.}(2019)Zonca, Singer, Lenz, Reinecke, Rosset, Hivon, \&
  Gorski}]{Zonca2019}
Zonca, A., Singer, L., Lenz, D., {et~al.} 2019, Journal of Open Source
  Software, 4, 1298

\bibitem[{Zonca {et~al.}(2021)Zonca, Thorne, Krachmalnicoff, \&
  Borrill}]{pysm3}
Zonca, A., Thorne, B., Krachmalnicoff, N., \& Borrill, J. 2021, Journal of Open
  Source Software, 6, 3783

\end{thebibliography}
